\documentclass[a4paper,11pt]{article}
\usepackage{jheppub}

\usepackage{amssymb}
\usepackage{amsmath}
\usepackage{shuffle}
\usepackage{graphicx}
\usepackage[colorlinks=true]{hyperref}
\usepackage[dvipsnames]{xcolor}
\usepackage{array}
\usepackage{tikz}
\usepackage{amsmath}
\usepackage{mathrsfs}
\usepackage{enumitem}
\usetikzlibrary{arrows.meta}
\usepackage{graphicx}
\usepackage{xspace}
\usepackage[dvipsnames]{xcolor}
\usepackage{pifont}
\usepackage{makecell}
\usepackage{multirow}
\usepackage{changepage}
\usepackage{pifont}
\usepackage{stackengine}
\usepackage{capt-of}
\usepackage[super]{nth}
\usepackage{braket}
\usepackage{cleveref}
\usepackage{subcaption}
\usepackage{xcolor}
\usepackage{bm}

\newcommand{\xmark}{\ding{55}}
\def\e{\epsilon}
\newcommand{\code}[1]{\texttt{#1}}
\newcommand*{\rom}[1]{\expandafter\@slowromancap\romannumeral #1@}

\newcommand{\liverpool}{Department of Mathematical Sciences, University of Liverpool, Liverpool L69 3BX, 
U.K.
}

\DeclareMathOperator{\rank}{rank}
\DeclareMathOperator{\Tr}{Tr}
\newcommand{\cmark}{\ding{51}}%

\allowdisplaybreaks

\preprint{{\raggedleft
            TUM-HEP 1529/24\\
		MPP-2024-193\\
            ZU-TH 52/24\\
            
}}

\title{Graded transcendental functions: an application to four-point amplitudes with one off-shell leg}

\author[a]{Thomas Gehrmann,}
\author[b]{Johannes Henn,}
\author[a]{Petr Jakub\v{c}\'{i}k,}
\author[b]{Jungwon Lim,}
\author[c]{Cesare Carlo Mella,}
\author[c]{Nikolaos Syrrakos,}
\author[c]{Lorenzo Tancredi,}
\author[d]{and William J. Torres~Bobadilla}

\affiliation[a]{Physik-Institut, Universit\"{a}t Zurich, 
            Winterthurerstrasse 190,
           	CH-8057 Z\"{u}rich,
            Switzerland}
\affiliation[b]{Max-Planck-Institut für Physik, Werner-Heisenberg-Institut, Boltzmannstr. 8, 85748 Garching,
Germany}
\affiliation[c]{Technical University of Munich, TUM School of Natural Sciences, Physics Department, James-Franck-Straße 1, 85748 Garching, Germany}
\affiliation[d]{\liverpool}
            
\emailAdd{thomas.gehrmann@physik.uzh.ch}
\emailAdd{henn@mpp.mpg.de}
\emailAdd{petr.jakubcik@physik.uzh.ch}
\emailAdd{wonlim@mpp.mpg.de}
\emailAdd{cesarecarlo.mella@tum.de}
\emailAdd{nikolaos.syrrakos@tum.de}
\emailAdd{lorenzo.tancredi@tum.de}
\emailAdd{torres@liverpool.ac.uk}

\abstract{ 
Several 
recent works have demonstrated the   powerful 
algebraic simplifications that can be 
achieved for scattering amplitudes through 
a systematic grading of transcendental quantities. 
We develop these concepts to construct a minimal basis of functions 
tailored to a scattering amplitude in a general way.
Starting with formal solutions for all master integral topologies, 
we organise the appearing functions by properties such as their symbol alphabet or letter adjacency.
We rotate the basis such that functions with spurious features appear 
in the least possible number of basis elements. 
Since their coefficients must vanish for physical quantities, this approach avoids complex cancellations.
As~a~first application, we evaluate all integral topologies relevant to
the three-loop $Hggg$ and $Hgq\bar{q}$ amplitudes in the leading-colour
approximation and heavy-top limit. 
We describe the derivation of canonical differential equation systems
and present a method for fixing boundary conditions
without the need for a full functional representation.
Using multiple numerical reductions, we test the maximal transcendentality conjecture for $Hggg$ and identify a new letter which appears in functions of weight 4 and 5.
In addition, we provide the first direct analytic computation of a three-point form factor of the operator $\Tr(\phi^2)$ in planar $\mathcal{N}=4$ sYM and find agreement with numerical and bootstrapped results.}

\begin{document}
       
\maketitle

\section{Introduction}
\label{sec: introduction}
The study of Feynman integrals has flourished in recent decades.
Particle phenomenologists view them as building blocks of multi-loop scattering amplitudes instrumental in providing higher-order predictions to match the precision programme of current colliders.
For theorists, they represent a rich but well-defined set of problems on the intersection of several mathematical fields from algebraic geometry to functional analysis.
Modern amplitude methods often drastically simplify the derivation of integrands, or provide recursion relations in highly symmetric theories.
Nonetheless, in realistic theories and with more than one loop, they invariably resort to the usual evaluation of Feynman integrals.

For a long time, linear identities such as integration-by-parts (IBP)~\cite{Chetyrkin:1981qh, Laporta:2000dsw} and Lorentz identities (LI) have been exploited to relate all integrals in a given problem to a smaller number of master integrals.
Master integrals are mostly solved as a Laurent expansion in the dimensional regulator $\e = (4-d)/2$ using the method of differential equations (DEs)~\cite{Kotikov:1991pm, Bern:1993kr, Remiddi:1997ny, Gehrmann:1999as}.
The evaluation of the relevant master integrals often precedes the computation of the amplitude because the tensor rank of the integrals in the amplitude can be high, implying a large set of identities with complicated rational coefficients.
Moreover, a single integral topology appears in many interesting physical quantities and in various theories.
This disconnect between the definition of the relevant transcendental functions, and the evaluation of their rational coefficients in a particular amplitude can lead to several issues.
Certain analytic structures repeat in the master integrals, possibly including functions which have unphysical branch cuts or terms which eventually drop out of the amplitude in $d=4$.
Such redundancy greatly contributes to the swell in the size of expressions at intermediate stages of amplitude calculations.
An approach which allows to make the set of transcendental functions conform with an amplitude at the stage when the integrand or reduction are not necessarily known, solely from constraints or conjectures about the final physical result, would be highly desirable. 

The idea of \emph{grading} transcendental functions to simplify the structure of physical 
amplitudes has a long history. 
Leveraging the symbol map~\cite{Goncharov:2010jf,Duhr:2011zq}, the first ideas in this direction were put forward for multiple polylogarithms more than a decade ago, especially in the context of the calculation of form factors in supersymmetric theories and closely related $H$+jet amplitudes in QCD~\cite{Brandhuber:2012vm,Duhr:2012fh}.
More recently, these concepts have been systematised, in combination with the method of canonical differential equations~\cite{Henn:2013pwa}, to grade the space of functions relevant for the calculation of various classes of five-point scattering amplitudes~\cite{Gehrmann:2018yef,Chicherin:2020oor,Chicherin:2021dyp}. 
In these references, the grading has been successfully used to reduce the appearance of spurious letters and to efficiently organise the numerical evaluation of the ensuing \emph{pentagon functions}. These techniques were 
applied subsequently to other 
 classes of two-loop four- and five-point 
integrals~\cite{Badger:2021nhg,Badger:2023xtl,Abreu:2023rco}.
In these works, the starting point is 
a formal solution for the master integrals in terms of Chen iterated integrals~\cite{Chen:1977oja},
which then allows to organise the space of functions into different classes of analytic complexity, including the appearance of specific letters. 
The members of each class are linearly independent 
when truncated at a given 
order, which prevents eventual cancellations and leads to an efficient representation of the amplitude. 

In this paper, 
we describe these developments in a general framework and extend the method in various aspects, such as the 
fulfilment of adjacency conditions or leveraging the grading of the functions to put constraints on the finite remainder of the amplitude using only the prediction for their $\e$-poles.
Firstly, we are able to completely exclude several types of functions from appearing in any amplitude relying on a given set of integrals, in any gauge theory.
Secondly, we show how the unique combinations of iterated integrals of different weights can be used to put constraints on the rational functions which multiply the transcendental functions in the finite part, even before the integrand or its reduction are known.
Finally, we demonstrate how to fully exploit the simplicity of
supersymmetric theories and the maximal transcendentality principle~\cite{Kotikov:2001sc} to effectively project QCD amplitudes onto their supersymmetric counterpart, plus a potentially weight-suppressed remainder.
Thanks to these improvements on the general method, we were able to determine the new functions appearing in the leading-colour $H$+jet amplitudes at three loops, bypassing full reduction, and write this most complicated part of the finite remainder in terms of a handful of functions with very simple rational coefficients.

Additionally, we describe a method for the determination of boundary constants suitable in a setting where many functions present in the master integrals are suspected to drop out of the amplitude.
Specifically, we avoid committing to a functional representation and an investigation of its properties, such as analytic continuation and numerical evaluation.
Instead, we identify singular one-dimensional kinematic slices where a solution in terms of multiple polylogarithms (MPLs)~\cite{Goncharov:1998kja, Remiddi:1999ew} in one variable is obtained and the expected asymptotic behaviour of the master integrals close to various singular surfaces can be easily enforced.
Through a matching procedure on the intersections of these slices, we transport all the information to a single point and completely fix the value of all the integrals in terms of a few single-scale integrals. 

These techniques are particularly suited for the computation of three-loop four-point Feynman integrals with massless propagators and one off-shell leg. 
As of today, analytic results for the corresponding two-loop integrals appeared 
in~\cite{Gehrmann:2000zt, Gehrmann:2001ck} in terms of 2dHPLs, which form a subset of MPLs.
More recently, these results were extended to higher transcendental weight~\cite{Gehrmann:2023etk} and applied to the computation of two-loop scattering amplitudes for $H$+jet production and the production of a vector boson $V=\gamma^*,Z,W^{\pm}$ and a jet~\cite{Gehrmann:2023zpz}.
At three loops, results for the simplest ladder topology first appeared in~\cite{DiVita:2014pza}, while the complete set of planar three-loop integrals, including tennis court topologies, was derived in~\cite{Canko:2021xmn}.
The planar integrals were recently used for the computation of the leading colour three-loop corrections for $V$+jet production~\cite{Gehrmann:2023jyv}.
First results on a few non-planar topologies appeared in~\cite{Henn:2023vbd} and subsequently in~\cite{Syrrakos:2023mor}.

The set of Feynman integrals considered in this paper covers the leading-colour contribution to the three-loop scattering amplitude for $H+$jet production in the infinite top-quark mass limit. 
Here, non-planar Feynman integrals contribute even at leading colour, unlike in the $V$+jet amplitude, where non-planar integrals appear only in the colour-suppressed part.
Evaluating all the master integrals for this process is one of the main results of this work. 
Despite their relative simplicity, the leading-colour terms are expected to dominate the contribution of the amplitude to a cross-section, and will be tackled first in the future. 
Higgs-plus-jet production and the closely related Higgs transverse momentum distribution are observables that allow precision studies of the Higgs 
sector of the Standard Model. 
Given that state-of-the-art second-order QCD predictions~\cite{Boughezal:2015aha, Boughezal:2015dra, Caola:2015wna, Chen:2016zka} have an estimated uncertainty of $\mathcal{O}(5\%)$, an extension to the next order is mandatory ahead of the high-luminosity run at the LHC. 
The amplitude is likewise an ingredient for the calculation of differential observables for Higgs production in gluon fusion at N$^4$LO.

First attempts at the non-planar topologies~\cite{Henn:2023vbd, Syrrakos:2023mor} revealed an extended symbol alphabet in the three-loop master integrals compared to previous loop orders. 
Essential to the philosophy of constructing graded transcendental functions is that insights from idealised theories can restrict the master functions, especially in a setting like the $H$+jet amplitude, where a full reduction of the amplitude is presently unfeasible. 
In this context, the maximally supersymmetric Yang-Mills theory ($\mathcal{N}=4$ sYM) plays a special role as it shares aspects of QCD but its high degree of symmetry allows computation to very high loop orders.

Especially valuable are the $\mathcal{N}=4$ \textit{form factors}: matrix
elements of a gauge-invariant composite operator between the vacuum and some
states of the theory\footnote{Following standard notation, $\phi_{ij}$ are the 6 scalar fields of the theory with $i,j=1,2,3,4$ and $\phi_{ij} = -\phi_{ji}$.}. 
As objects situated at the interface between on-shell quantities like scattering amplitudes and off-shell quantities such as correlation functions, they interpolate between the disparate tools used to compute them. 
Particularly important is the operator $\Tr(\phi^2_{12})$ which has vanishing anomalous dimension, i.e. it is half-BPS protected, and its form factors seem to obey the \textit{maximum transcendentality principle}~\cite{Kotikov:2002ab, Kotikov:2004er}. 
According to this principle, the maximally transcendental part of the finite remainder of specific quantities in QCD coincides with the equivalent form factor in $\mathcal{N}=4$ sYM. 
The simplest case, described by universal cusp and collinear dimensions, is the Sudakov form factor with two scalar fields, where the conjecture has been verified by direct computation at two~\cite{vanNeerven:1985ja}, three~\cite{Gehrmann:2011xn, Gehrmann:2011aa} and four~\cite{Lee:2021lkc, Lee:2022nhh} loops.

Remarkably, this property was also observed between the $Hggg$ amplitude in Higgs Effective Field Theory (HEFT) and the form factor involving two scalars and a single gluon with positive helicity up to two loops~\cite{Brandhuber:2012vm}, processes with non-trivial kinematics. 
At three loops, the integrand of this quantity was derived using colour-kinematics duality and unitarity cuts~\cite{Lin:2021kht,Lin:2021qol}. 
It has been evaluated numerically firstly using \code{FIESTA}~\cite{Smirnov:2021rhf} and \code{pySecDec}~\cite{Borowka:2017idc}, and subsequently at higher precision with \code{AMFlow}~\cite{Liu:2017jxz, Liu:2022chg} in~\cite{Guan:2023gsz}. 
Using the symbol bootstrap, the analytic result is known at the symbol level through 8 loops~\cite{Dixon:2020bbt,Dixon:2022rse}. 
All of these results show exceptional simplicity and contain only integration kernels familiar from the one- and two-loop amplitudes. 
Here, we evaluate the planar part of the $\Tr(\phi^2_{12})$ form factor for the first time with a direct analytic calculation.
Using several numerical reductions of the three-loop $Hggg$ amplitude, we can verify that the maximally transcendental part shares the simple alphabet of the form factor. 
This result is a springboard for tailoring the space of functions for the QCD amplitude.

The remainder of the paper is structured as follows. 
In section~\ref{sec: functions}, we describe the construction of graded transcendental functions. 
The notation and conventions for the $H$+jet amplitude and $\Tr(\phi^2)$ form factor are introduced in section~\ref{sec: canonical}, where we also elaborate on the derivation of canonical bases for all integral families. 
The fixing of boundary conditions is then described in section \ref{sec: boundaries}. 
In section~\ref{sec: results} we present the result for the form factor, observations on the analytic structure of the $H$+jet amplitude, and perform several checks before drawing conclusions in section~\ref{sec: conclusion}. 
\section{Graded transcendental functions}
\label{sec: functions}
Traditionally, the computation of master integrals for an amplitude amounts to listing the relevant topologies, finding a set of Feynman integrals independent in $d=4-2\e$ dimensions under IBP, Lorentz and symmetry identities, and writing for them a Laurent expansion in $\e$ up to the highest order which can appear in the finite part of the amplitude. 
This approach is certainly sufficient, but often overly complicated. 
It is common knowledge that master integrals might contribute to physical quantities in $d=4$ dimensions only in particular combinations. 
In addition, there are master integrals whose coefficients are explicitly suppressed by one or more powers of $\e$.
As a consequence, it might happen that some of the analytic structures appearing in the individual master integrals (for example specific types of branch cuts in their solution) do not show up in the finite part of a scattering amplitude.
This is typically  observed \emph{a posteriori} through an involved pattern of cancellations among different integrals which is difficult to foresee.
Making such cancellations manifest before the insertion of all integrals into the amplitude would clearly provide a substantial improvement to the standard approach.

In this section, we review and extend a strategy to achieve this goal, building upon the original 
developments that were made in the 
context of pentagon functions~\cite{Chicherin:2020oor,Chicherin:2021dyp}.
The principal idea is to consider a \textit{formal} solution for the corresponding
Feynman integral topologies to find the minimal set of functions without spurious or evanescent features with regard to a physical amplitude. 
In particular, we algorithmically determine a genuine basis of functions that
are independent under all linear identities and divided into distinct subsets according to their analytic properties.
As a consequence, parts of these functions might never have to be evaluated or solved explicitly beyond their formal representation. 

\subsection{Vector space of transcendental functions}
The Feynman integrals appearing in a scattering amplitude are related through linear equations to a small number of master integrals (MIs). 
In the standard approach, one derives a coupled set of differential equations (DEs) for the master integrals and determines their exact numerical 
values at a kinematic point. 
Formally, the DEs and the boundary constants constitute the solution. 
As a Laurent series in the dimensional regularisation parameter $\e$, the latter can typically be expressed in terms of iterated integrals with kernels which stem from the entries of the DE. 
This form becomes especially transparent when the DEs are
$\e$-factorized~\cite{Henn:2013pwa}. 
Their solution can then be expressed in terms of 
Chen iterated integrals~\cite{Chen:1977oja}
\begin{align}
I(\omega_1,\ldots,\omega_n;\vec{x}) = 
\int_{\gamma}\omega_1\,  \omega_2 \cdots \omega_n, 
\qquad I(;\vec{x}) = 1\,,
\label{eqs:Chen_int}
\end{align}
where $\omega_i = \omega_i(\vec{x})$
are (linearly independent) differential forms in the 
kinematic invariants  
and $\gamma = \gamma(\vec{1}_0,\vec{x})$ is a curve connecting the base point 
$\vec{1}_0$ to a generic kinematic point $\vec{x}$. 
The formal representation becomes explicit after parametrising  
the differential forms along the path $\gamma$
\begin{align}
\int_{\gamma}\omega_1  \cdots \omega_n = \int_{t_0}^t dt_n f_n(t_n)
\int_{t_0}^{t_n} dt_{n-1} f_{n-1}(t_{n-1}) \cdots
\int_{t_0}^{t_2} dt_{1} f_{1}(t_{1})\,, \label{eq:Chen_int_f}
\end{align}
where $\gamma(t) = \vec{x}$ with $\gamma(t_0) = \vec{1}_0$, and 
the complex functions $f_i(t_i)$ are the pull-backs of the differential forms
along the path, $\gamma^* \omega_i = dt_i f_i(t_i)$.
The set of differential forms $\{\omega_i\}$, known as the \textit{alphabet}, is made up of individual \textit{letters} and it can be read off directly from the differential equations in $\e$-factorised form. 
For a general set of differential forms, this remains a formal solution unless one can fully describe all functional identities among the forms and compute series expansions close to singular and regular points. 
This is required to impose the boundary conditions which
uniquely determine the solution as well as for numerical evaluation
of the iterated integrals.

If we limit ourselves to the case of $d\log$ forms, which will be relevant in this paper, more structure can be given to the iterated integrals.
In particular, the number of successive integrations $n$, or \textit{length} of the iterated integral, can be identified with the \textit{transcendental weight} of the corresponding function.
The same definition can be extended to transcendental numbers $\xi_n = \pi^{n}, \zeta_n,\ldots$, which correspond to special values of the iterated integrals, and we also assign weight $-1$ to $\e$. 
Additionally, we call an expression \textit{uniformly transcendental} (UT) if it is a sum of terms of the same weight and \textit{pure} if the weight is lowered by one by taking a derivative. 
In practice, pure functions can only contain iterated integrals with rational numbers as prefactors, not rational functions.

Iterated integrals feature a weight-preserving product formula
\begin{align}
I(\vec{\omega}_\alpha;\vec{x})I(\vec{\omega}_\beta;\vec{x})
=\sum_{\vec{\omega}=\vec{\omega}_\alpha \shuffle\, \vec{\omega}_\beta}I(\vec{\omega};\vec{x})\,,
\label{eqs: shuffle product}
\end{align}
where the \textit{shuffle} $\shuffle$ denotes all order-preserving permutations of the lists of differential forms $\vec{\omega}_\alpha, \vec{\omega}_\beta$. 
For a thorough introduction to Chen iterated integrals and their use in multiloop computations, see also~\cite{Zoia:2021zmb}.
If we apply~\eqref{eqs: shuffle product} a maximal number of times, we arrive at a form which has no hidden zeros, i.e.\ where two expressions involving functions $I$ with different sets of differential forms are genuinely different. 
We assume that the transcendental constants which appear in the solutions for MIs are likewise written in an independent way.
Thanks to this property, products of the shuffled functions $I$, powers of transcendental numbers, and the dimensional regulator
\begin{align}
b_{\mathcal{T}_w}=\{\e^{-a}\xi_{n}^{b}I(\omega_1,\ldots,\omega_c;\vec{x})\}\,,\qquad w=a+nb+c
\label{eq:defTw}
\end{align}
form a basis of the vector space $\mathcal{T}_w$, which we define to contain all transcendental functions \textit{up to} weight $w$ with algebraic numbers as coefficients. 

Scattering amplitudes truncated at an $\e$ order containing functions of weight at most $w$ are combinations of some set of transcendental functions 
with algebraic functions as prefactors,
\begin{align}
    A_{w} =\sum a(\vec{x}) \cdot b_{\mathcal{T}_w}\,.
\end{align}
At a fixed but arbitrary phase-space point\footnote{
We fix the algebraic functions to a numerical point (as coordinates) and keep the transcendental functions symbolic (as basis vectors).}, 
amplitudes (as well as Feynman integrals truncated at weight $w$) are therefore vectors in the space $\mathcal{T}_w$.
As an instance of the $n$-dimensional complex vector space of algebraic numbers $\mathbb{A}^{n}$, the space is equipped with the standard dot product. 

The choice of the set of functions $b_{\mathcal{T}_w}$ has an influence on the complexity of the algebraic coefficients $a(\vec{x})$. 
For example, redundancy in the transcendental functions might lead to the presence of many hidden zeros in the algebraic coefficients. 
In the following, we use the above construction to determine a basis which is minimal when truncated at a given $\e$-order and manifests the expected physical properties of the amplitude. 
We do not require explicit knowledge of the integrand, nor the Feynman rules or details of the theory, only a formal solution for a (not necessarily minimal) set of integrals which describe it.

\subsection{Finding the minimal subspace}
\begin{figure}[t]
\includegraphics[width=\textwidth]{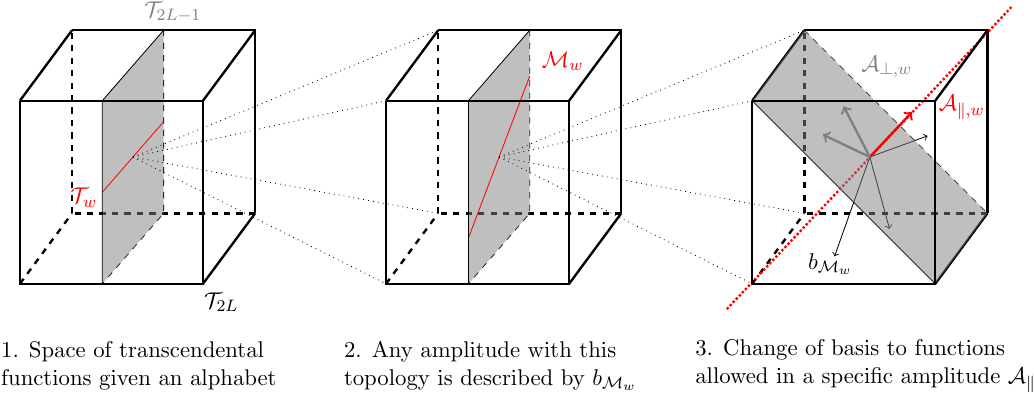}
\vspace*{-4mm}
\caption{Illustration of the construction of a basis of functions tailored to a scattering amplitude. 
At any phase-space point, an amplitude resides in the vector space $\mathcal{T}_w$ over all elemental transcendental functions with the relevant alphabet up to weight $w$ (left figure). 
At any fixed weight (middle figure), only some of the functions appear in the relevant integral topologies, restricted by their DEs and boundary conditions (grey surface). 
Due to repeated patterns and linear combinations across master integrals, the actual space containing the formal solutions for the MIs is typically smaller (red line). 
It is spanned by the minimal set of functions required to represent any amplitude with these topologies. 
For a specific amplitude (right figure), conventional master integrals (black vectors) can be replaced by a basis of functions adhering to any restrictions on the amplitude (red vector), and unphysical combinations of functions in the orthogonal complement (grey vectors) which must drop out.
}
\label{Fig2: three cubes}
\end{figure}
Starting with the vector space $\mathcal{T}_{w}$ of transcendental functions up to weight $w$, we will progressively curtail the subspace in which the amplitude and hence the ideal basis of functions can reside. 
The space spanned by all possible iterated integrals generated with all permutations of the differential forms appearing in the DEs 
for a given problem is clearly too large. 
Indeed, DEs together with boundary conditions immediately forbid
some combinations, in other words, some coordinates in $\mathcal{T}_{w}$ are always zero. 
In the remaining coordinates, the master integrals span a relatively small
subspace due to relations between the coefficients of the iterated integrals 
in the formal solutions. Firstly, the iterated nature of the solution implies
that the same patterns within the expansion for a single MI repeat at higher $\e$ orders. Secondly, individual integrals might be symmetric in some kinematic variables and thus the coefficients of the elemental functions $I$ with different arguments coincide. Lastly, there can be genuine symmetries between integrals from various integral families, their kinematic crossings, or ones that emerge accidentally due to the truncation. 

All of these patterns and identities can systematically be taken into account (irrespective of their origin or interpretation) with standard linear algebra, 
for instance by row-reducing the matrix $S$ formed by the coordinates of the formal solutions $F_i$ of the $n_{\mathrm{MI}}$ master 
integrals with respect to the elements of $b_{\mathcal{T}_{w}}$, 
\begin{align}
    S_{ij} = \left[F_i\right]_{b_{\mathcal{T}_{w},j}}\quad\quad i = 1,\dots,n_{\mathrm{MI}}; \,\, j = 1,\dots,\dim(\mathcal{T}_{w}) \,.
    \label{eq: Smatrix}
\end{align}
The subspace $\mathcal{M}_w \subseteq \mathcal{T}_w $ spanned by the master integrals (the row space of $S$) 
is the minimal set of functions which suffice to express \textit{any} quantity
which depends on the integral topologies in question. 
Crucially, the rank of $S$ is the dimension of both its row space and its column space. 
In nearly all amplitude computations, the number of master integrals (number of rows of $S$) is significantly lower than the number of elemental transcendental functions $I$ up to a certain weight (number of columns of $S$),
\begin{align}
    \dim(\mathcal{M}_w) = \rank(S) \leq n_{\mathrm{MI}} \ll \dim(\mathcal{T}_{w})\,.
\end{align}
In plain language, many iterated integrals $I$ with different permutations of differential forms might appear in the integrals, but the number of their linearly independent combinations is bounded by the number of MIs and often even
smaller. 
Identifying the row space of the matrix $S$ therefore leads to a
significantly more economical basis than the MIs or, even worse, all individual
iterated integrals. 
In practice, the transformation from the original basis of master integrals to the minimal set of functions can be found by row-reducing the matrix $S$ and solving for the original master integrals in terms of the independent rows.
If we also find all the linear combinations of the original masters which are zero up to this weight (a basis of the left nullspace of $S$), we can write the full transformation $T'$ as an invertible matrix over algebraic numbers,
\begin{align}
    \vec{J}_w = T'\cdot 
    \begin{pmatrix}
        b_{\mathcal{M}_{w}}\\
        \vec{0}_w
    \end{pmatrix}\,,
\end{align}
which expresses the original master integrals $\vec{J}_w$ truncated at weight $w$ in terms of the minimal basis.

Furthermore, a \textit{specific} physical scattering amplitude occupies a yet smaller subspace of $\mathcal{M}_w$, defined, e.g. by the absence of certain letters, their occurrence only at some weights, restrictions on their adjacency, or behaviour in unresolved limits. A further rotation on the functions in $\mathcal{M}_w$ can therefore be applied such that the elements are categorically in the physically relevant subspace $\mathcal{A}_{\parallel,w}$, or in the orthogonal complement $\mathcal{A}_{\perp,w}$:
\begin{align}
\mathcal{M}_w = \mathcal{A}_{\parallel,w} \oplus \mathcal{A}_{\perp,w}\,.
\end{align}
Note that the original master integrals have in general non-trivial 
projections onto both of the subspaces. After rotation, the
functions in $\mathcal{A}_{\perp,w}$ must be simply absent in the
amplitude (their coefficient $0$), avoiding spurious cancelations when
expressing the amplitude in this basis. The same construction can also be
applied  whenever we want to organise the appearing functions into
subsets which should not mix, introducing a \textit{grading}. 

In the common case where the enforced property is the absence of certain functions~$I$, we simply remove the corresponding coordinates (columns in $S$), find the new, potentially smaller row space, and map the integrals in $b_{\mathcal{M}_w}$ to it by solving a set of linear equations.
More generally, one can first find a set of independent vectors that span the desired subspace $\mathcal{A}_{\parallel,w}$ and a set of vectors that have a vanishing dot product with all elements of $b_{\mathcal{A}_{\parallel,w}}$, i.e. a basis for the orthogonal complement $\mathcal{A}_{\perp,w}$, and find the mapping to this new basis.
Often, we perform this transformation for functions truncated at a certain weight and subsequently extend the solutions to weight $w+1$, where the vectors whose projection onto ${\mathcal{M}_w}$ is entirely contained in $\mathcal{A}_{\parallel,w}$ can again be split into orthogonal subspaces relevant and irrelevant to the amplitude at this weight, without affecting $\mathcal{A}_{\perp,w}$.

The additional rotations within the elements of $b_{\mathcal{M}_w}$ can again be summarised in an invertible transformation matrix $T''$ with size $n_{\mathcal{M}} = \dim(\mathcal{M}_w)$.
As a result, we obtain the final mapping from the original master integrals $\vec{J}_w$ to the $n_{\psi}=\dim(\mathcal{A}_{\parallel, w})$ minimal tailored functions $\vec{\psi}_w$: 
\begin{align}
\begin{pmatrix}
    \vec{\psi}_w\\
    \vec{0}_w
\end{pmatrix}
&= 
\begin{pmatrix}
    T'' & \phantom{T''} \\
    & \mathbb{I}
\end{pmatrix}
\cdot
(T')^{-1}
\cdot
\vec{J}_w \nonumber\\
&\equiv 
T \cdot \vec{J}_w\,,
\end{align}
where we indicated the irrelevant functions which are zero up to weight $w$ with $\vec{0}_w$.
This transformation can also be applied to the original system of DEs for the master integrals in the standard way,
\begin{align}
\mathrm{d}\vec{J}(\vec{x}; \e) & = \mathrm{d}M(\omega_i;\e) \,\vec{J}(\vec{x};\e)\,,\\
    \mathrm{d}
    \begin{pmatrix}
        \vec{\psi}_w(\vec{x};\e) \\
        \vec{0}_w
    \end{pmatrix} &= [T\cdot \mathrm{d}M(\omega_i;\e) \cdot T^{-1}]\cdot  \begin{pmatrix}
        \vec{\psi}_w(\vec{x};\e) \\
        \vec{0}_w
    \end{pmatrix}\,.
    \label{eq: rotated DEs}
\end{align}
The first $n_\mathcal{M}$ rows and columns of eq.~\eqref{eq: rotated DEs} represent a closed set of differential equations for the minimal tailored functions, valid up to weight $2L$. 
Additionally, the transformation produces $n_{\mathrm{MI}} - n_{\mathcal{M}}$ non-trivial differential equations whose solution is, however, by construction vanishing up to weight $2L$ and which can be simply discarded. 
Moreover, this transformation preserves potential $\e$-factorisation of the matrix $\mathrm{d}M$.
Solving the rotated system in terms of an explicit representation of the transcendental functions (instead of the original DEs) might be advantageous because the system is smaller and contains the least possible number of elements with the undesired properties which were the target of the rotations, such as problematic alphabet letters.
Fig.~\ref{Fig2: three cubes} provides a pictorial summary of our method.

\subsection{Propagating information from poles to the finite remainder}
\label{subsec: poles to finite}
In this subsection, we will focus on master integrals which are UT and their Laurent expansion is normalised to start at $\mathcal{O}(\e^0)$.
To allow also for amplitudes which mix terms of different weights,
it is equivalent and convenient to work with coefficients $a(\vec{x};\e)$ which have an $\e$ expansion, rather than considering multiple $\e$ powers for each iterated integral in~$b_{\mathcal{T}_w}$, as in~\eqref{eq:defTw}.
Conditions on the non-appearance of certain transcendental functions will simply be translated to the vanishing of certain $\e$ orders in the expansion of the algebraic prefactor.

Generally in gauge theories, the ultraviolet (UV) and infrared (IR) poles of an amplitude in dimensional regularisation, i.e. the first at most $2L$ orders in $\e$, are predicted by products of lower-loop amplitudes with the same external states and by universal constants, cusp and collinear anomalous dimensions.
It follows that if some feature was not present in the $L-n$-loop amplitude expanded to
$\mathcal{O}(\e^{2n-1})$, it cannot appear in the poles of the $L$-loop amplitude. 
As an example of such a feature, we will discuss the presence of a new alphabet letter $\bm{\omega_{\mathrm{new}}}$.

Despite this restriction on the amplitude, the master integrals for the $L$-loop process might feature new alphabet letters in iterated integrals at orders less than $\e^{2L}$.
In this case, we expect the coefficients of the corresponding MIs to conspire such that the iterated integrals with $\bm{\omega_{\mathrm{new}}}$ drop out of the poles of the amplitude.
With our method, we can identify the new iterated integrals at $w<2L$ as those residing in $\mathcal{A}_{w,\perp}$ 
and we can assign them to a minimal number of functions where their coefficients are linearly independent.
Then, any function containing a weight-$w$ iterated integral $I(\ldots,\bm{\omega_{\mathrm{new}}},\ldots)$ must be multiplied by a coefficient whose expansion starts at $\mathcal{O}(\e^{2L-w})$.
The weight-$w$ part of this function will appear in the finite part of the amplitude and the terms of weight $w+1$ and beyond become irrelevant.
Consequently, if an iterated integral at weight $w>w_0$ is preceded in every independent master 
function in which it occurs by an iterated integral of weight $w\leq w_0$ with a letter (or other feature) 
appearing at this loop order for the first time, it cannot survive in any 4-dimensional amplitude.
The power of this statement lies in the ability to use the knowledge of the minimal basis functions to restrict complicated iterated integrals or analytic structures from appearing in any physical quantity relying on a given graph topology.
For instance, an iterated integral with two or more integrations over some new kernel can only survive if it appears in a minimal function with no new kernels at lower weights.

These observations did not require explicit knowledge of the pole prediction or lower-loop amplitudes, only information about the presence of absence of certain features.
On top of this, most of the time, we can derive the full pole prediction for an amplitude and we can use it alongside the formal $L$-loop solutions to constrain the finite part of the $L$-loop amplitude without even knowing its integrand.
Firstly, the iterated integrals in the pole prediction must (after applying the shuffle product) always be a subset of those present in the $L$-loop master functions.
Solving for the pole prediction using the master functions, we can leverage the unique linear combinations which appear in them to restrict the coefficients of the iterated integrals also in the finite part.
In a basis of uniform transcendental weight, every weight-$w$ iterated integral can in this way yield up to $2L-w$ conditions on the $\e$-expanded coefficients of the minimal functions.
Generally speaking, the constraints are most powerful for functions with high transcendentality.
The weight-$2L$ functions in the finite part are constrained by the coefficients of all functions which appear in MIs up to weight $2L-1$.
On the other hand, for the single purely rational function in the finite part, this method does not provide any restrictions.

\paragraph{Example}
For clarity, we provide a fictional but instructive example in which we will constrain the finite part of a one-loop amplitude $A$ as a function of the kinematic invariants~$\vec{x}$.
Suppose the pole prediction for this amplitude is known and it contains logarithms of the invariants $\omega_1(\vec{x})$ and $\omega_2(\vec{x})$ in the single pole,
\begin{align}
    \mathscr{P}(A) = \frac{2}{\e^2} + \frac{-1+2I(\omega_1)+6I(\omega_2)}{\e}\,.
    \label{eq: ex poles}
\end{align}
Let us assume that this amplitude involves 3 canonical master integrals whose solutions are UT and normalised to start at $\mathcal{O}(\e^0)$,
\begin{alignat*}{4}
    \mathrm{MI}_1 &=  &&+ \e [2 I(\omega_1) + 3 I(\omega_2) + I(\bm{\omega_3})]  &&&+ \e^2[I(\omega_1,\omega_1) + I(\omega_2,\omega_1) - 4 I(\bm{\omega_3},\bm{\omega_3})] +\mathcal{O}(\e^3) \,,\\
    \mathrm{MI}_2 &= - 1 &&+ \e[I(\omega_1) + I(\bm{\omega_3})]  &&&+ \e^2[I(\omega_1,\omega_2) - I(\omega_2,\omega_1) - 4 I(\bm{\omega_3},\bm{\omega_3})]  +\mathcal{O}(\e^3)\,,\\
    \mathrm{MI}_3 &= -1 &&+\e[3I(\omega_1)  + 3 I(\omega_2)+ 2I(\bm{\omega_3})] &&&+ \e^2[I(\omega_1,\omega_1) + I(\omega_1,\omega_2) -8 I(\bm{\omega_3},\bm{\omega_3})]+\mathcal{O}(\e^3)\,.
    \label{eq: ex 3 MIs}
\end{alignat*}
Note that the MIs contain the letter $\bm{\omega_3}$, which is new at this loop order, already at $\mathcal{O}(\e^{1})$, potentially showing up in the simple pole of $A$. 
The space $\mathcal{T}_{w=2}$ is therefore spanned by all iterated integrals with kernels $\omega_1, \omega_2, \bm{\omega_3}$, powers of $\e$ and transcendental constants with total weight up to 2 in every term.
The MIs in~\eqref{eq: ex 3 MIs} are linearly independent combinations of integrals in $d$ dimensions, but their truncation to $\mathcal{O}(\e^2)$ introduces a degeneracy.
To remove it, we can apply the transformation
\begin{align}
    (T')^{-1} = \begin{pmatrix}
    1&0&0\\
    1&-1&0\\
    1&1&-1
    \end{pmatrix}\,
\end{align}
and obtain the minimal set of functions spanning $\mathcal{M}_{w=2}$,
\begin{alignat*}{4}
    \psi_1&=   &&+ \e [2 I(\omega_1) + 3 I(\omega_2) + I(\bm{\omega_3})]  &&&+\e^2[I(\omega_1,\omega_1) + I(\omega_2,\omega_1) - 4 I(\bm{\omega_3},\bm{\omega_3})] +\mathcal{O}(\e^3)\,,\\
    \psi_2&= 1 &&+\e[I(\omega_1)  + 3 I(\omega_2)] &&&+\e^2[I(\omega_1,\omega_1) - I(\omega_1,\omega_2) +2 I(\omega_2,\omega_1)]+\mathcal{O}(\e^3)\,
\end{alignat*}
and the third function is $\mathcal{O}(\e^3)$ and thus irrelevant.
Any physical quantity depending on these master integrals could be expressed in terms of the minimal set of functions $\psi_1, \psi_2$.

The coefficients of the minimal functions in the particular amplitude $A$ can be written as $\e$ expansions up to the last relevant order
\begin{align}
    c_{\psi_1}(\vec{x})&=\frac{1}{\e^2}c_{\psi_1}^{(-2)}(\vec{x})+\frac{1}{\e}c_{\psi_1}^{(-1)}(\vec{x})+c_{\psi_1}^{(0)}(\vec{x})\,,\\
    c_{\psi_2}(\vec{x})&=\frac{1}{\e^2}c_{\psi_2}^{(-2)}(\vec{x})+\frac{1}{\e}c_{\psi_2}^{(-1)}(\vec{x})+c_{\psi_2}^{(0)}(\vec{x})\,,
\end{align}
where the $c_{\psi_i}^{(j)}$ are rational functions of $\vec{x}$ only.
Immediately we can see that $c_{\psi_1}^{(-2)}(\vec{x}) = 0$ because $\psi_1$ is the only (and therefore linearly independent) function which contains an iterated integral not allowed in the poles, $I(\bm{\omega_3})$.
In other words, $\e I(\bm{\omega_3})\in \mathcal{A}_\perp$ and $\psi_1$ is the only function which has a non-zero projection onto the basis of this orthogonal subspace.
The vanishing of $c_{\psi_1}^{(-2)}(\vec{x})$ also implies that the entire finite part of $\psi_1$ may only appear beyond the finite part of the amplitude, and the function $I(\bm{\omega_3},\bm{\omega_3})$ is thus completely excluded from the amplitude.
To gain even more information, we can explicitly solve for some of the remaining coefficients by equating the expression for the amplitude in terms of the minimal functions with the poles in eq.~\eqref{eq: ex poles},
\begin{align}
\mathscr{P}(A)&\overset{!}{=} c_{\psi_1}(\vec{x}) \psi_1 + c_{\psi_2}(\vec{x}) \psi_2\,,
\end{align}
which allows us to fix also $c_{\psi_2}^{(-2)}$ and $c_{\psi_2}^{(-1)}$.
In summary, we learnt that
\begin{align}
    c_{\psi_1}^{(-2)}(\vec{x}) = 0\,, \quad c_{\psi_1}^{(0)}(\vec{x}) = \mathrm{irrel.}\,, \quad c_{\psi_2}^{(-2)}(\vec{x}) = 2\,, \quad c_{\psi_2}^{(-1)}(\vec{x}) = -1 
\end{align}
and the finite part of the one-loop amplitude has only two undetermined functions of the kinematics with no $\e$ dependence left,
\begin{align}
    \mathscr{F}(A) = &+ 2 I(\omega_1,\omega_1) - 2 I(\omega_1,\omega_2) + 4I(\omega_2,\omega_1)-I(\omega_1) - 3I(\omega_2)  \\
    &+ c_{\psi_1}^{(-1)}(\vec{x})[2I(\omega_1)+3I(\omega_2)+I(\bm{\omega_3})] + c_{\psi_2}^{(0)}(\vec{x}) + \mathcal{O}(\e^1)\,.
\end{align}

Three features of this example stand out.
Firstly, note that we had no knowledge of the one-loop amplitude or its integrand.
Secondly, some coefficients remain undetermined.
This will almost always be the case but the number of constraints is still significant.
Finally, if we decide to perform an IBP reduction and reconstruction of coefficients, the constraints help to significantly reduce the number of functions we need to determine.
Namely, the reconstruction of the $\e$-expanded coefficients of the original MIs would require determining 9 functions of $\vec{x}$.
Fitting the coefficients of individual iterated integrals leads to the reconstruction of 13 functions of $\vec{x}$, while only 2 functions need to be reconstructed to fit the coefficients of the proposed functions $\psi$.
The impact on the efficiency of the IBP reduction is therefore expected to be dramatic.
We have also seen that the appearance of the integrals $I(\bm{\omega_3})$ and $I(\bm{\omega_3},\bm{\omega_3})$ exclusively in $\psi_1$ meant that the former function is excluded from any amplitude which relies on these integrals.
Moreover, in Section~\ref{sec: results} we will demonstrate that in our problem, a significantly simplified IBP reduction can fix the remaining coefficient like $c_{\psi_1}^{(-1)}$ if the new function $I(\bm{\omega_3})$ which accompanies it stems from Feynman graphs with many internal lines.\\

In summary, 
building upon earlier works
made in the context of pentagon functions~\cite{Chicherin:2020oor,Chicherin:2021dyp}
we described and 
further developed a method which identifies a minimal set of functions under all linear relations among the formal solutions for the MIs relevant to some amplitude.
Next, we have shown how to split the functions into subsets of independent functions based on some property of the constituent iterated integrals.
This grading leads to an efficient organisation of the calculation, allows us to identify functions which conform to our physics expectations on the final amplitude and exclude the rest.
Even before the integrand of the amplitude is known, with this method, we can constrain the finite part of the end result and facilitate a future IBP reduction.
Finally, an amplitude expressed in this minimal and tailored set of transcendental functions is devoid of redundancy or cancellations, compact and manifestly obeying the properties which were enforced on the level of the basis of functions.
In the remainder of the paper, we will describe an application of this method to the computation of the three-loop correction to the form factor of $\Tr(\phi^2)$ in planar $\mathcal{N}=4$ sYM and to the helicity amplitudes for $H$+jet production at leading colour.
\section{Canonical integrals for four-point functions with one off-shell leg}
\label{sec: canonical}
We start with identifying the necessary integral topologies and deriving the corresponding 
canonical differential equations. The integrals for the decay of a massive particle 
with $q^2=m^2>0$ to three massless particles with momenta $p_1,p_2,p_3$ and $p_i^2=0$ are described by three dimensionless variables
\begin{align}
    x = \frac{2p_1\cdot p_2}{q^2}\,,\quad y = \frac{2p_1\cdot p_3}{q^2}\,,\quad z = \frac{2p_2\cdot p_3}{q^2}\,,
    \label{eqs:kinematic}
\end{align}
which obey momentum conservation
\begin{align}
    x+y+z=1\,.
\end{align}
The Mandelstam invariants are defined as $s_{ij}=2p_{i}\cdot p_j$. For simplicity, we work in the \textit{Euclidean} kinematic region defined by 
\begin{align}
    0<z<1\,,\quad 0<y<1-z\,,\quad x=1-y-z\,.
\end{align}
We will address the problem of the analytic continuation to all the channels of $2\to2$ \textit{scattering} in a future
publication.
Nevertheless, we will see below that focusing on the QCD amplitude rather than individual master integrals can simplify this task considerably.

In $\mathcal{N}=4$ super-Yang-Mills theory, 
the simplest object with this kinematics is the three-point 
form factor defined as the overlap of the state created by the half-BPS 
operator $\Tr(\phi_{12}^2)$ and a state formed by two scalars and a positive-helicity gluon,
\begin{align}
    F_{\Tr(\phi^2)}
    = \int d^dxe^{-iq\cdot x}\Braket{\phi_{12}^{a_1}(p_1)\phi_{12}^{a_2}(p_2)g_{+}^{a_3}(p_3)|\Tr(\phi_{12}^2)|0}\,,
    \label{eq: form factor definition}
\end{align}
where $\phi_{ij}$ ($i,j=1,2,3,4$ with $\phi_{ij} = -\phi_{ji}$) denote the 6 on-shell scalar superfields. 
The significance of $\Tr(\phi_{12}^2)$ lies in its relation to the stress-tensor supermultiplet which includes the stress-tensor and other conserved currents, as well as the Lagrangian of the theory. 
The form factor has a single independent helicity configuration ($\mathrm{MHV}$), with the $\mathrm{\overline{MHV}}$ configuration related by symmetries, for details see~\cite{Eden:2011yp, Brandhuber:2011tv, Bork:2014eqa}.
\begin{table}[t!]
\begin{center}
\resizebox{\textwidth}{!}{
\begin{tabular}{m{2.45cm}cm{2.45cm}lm{2.45cm}lm{2.45cm}l}
\multicolumn{1}{c}{\textbf{gluonic webs}} &&\multicolumn{6}{c}{\textbf{non-planar topologies for $H$+jet at l.c.}}\\[10pt]
\includegraphics[width=0.16\textwidth]{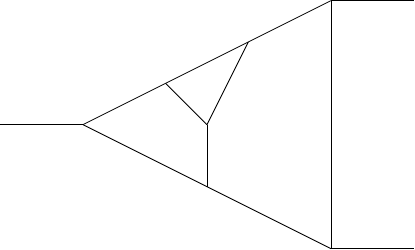}&$\Rightarrow$&\includegraphics[width=0.16\textwidth]{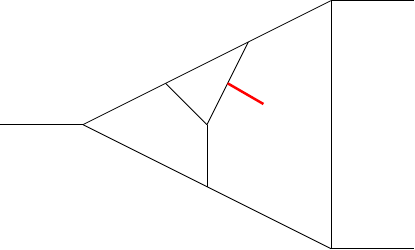}&$F_1$;&\includegraphics[width=0.16\textwidth]{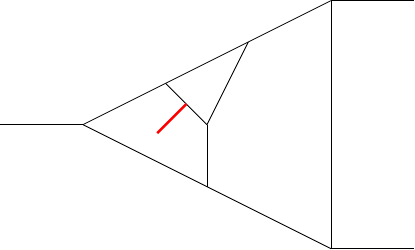}&$G_1$;&\includegraphics[width=0.16\textwidth]{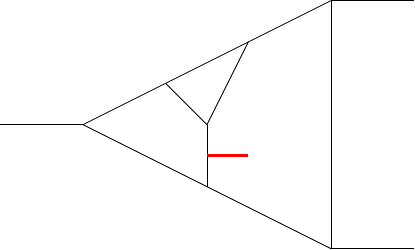}&$B_6^*$\\[20pt]
\includegraphics[width=0.16\textwidth]{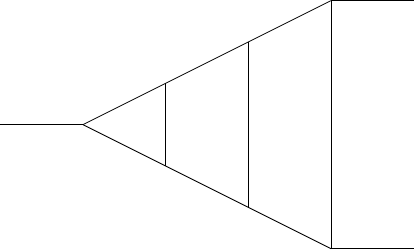}&$\Rightarrow$&\includegraphics[width=0.16\textwidth]{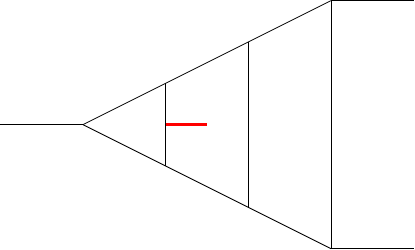}&$B_1$;&\includegraphics[width=0.16\textwidth]{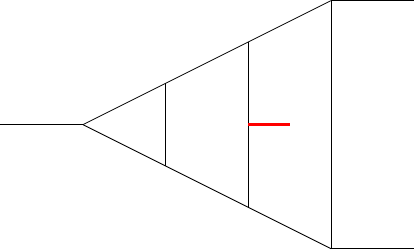}&$B_5^*$&&\\[20pt]
\includegraphics[width=0.16\textwidth]{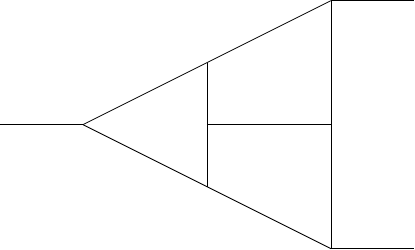}&$\Rightarrow$&&&\includegraphics[width=0.16\textwidth]{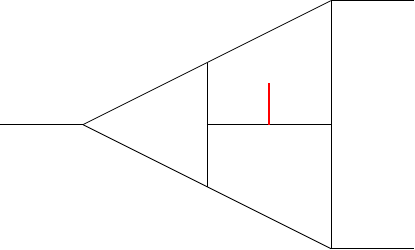}&$I_2$&&\\[20pt]
\includegraphics[width=0.16\textwidth]{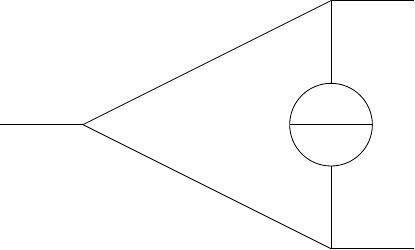}&$\Rightarrow$&&&\includegraphics[width=0.16\textwidth]{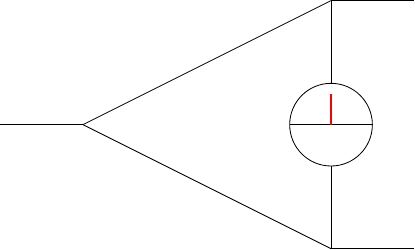}&$B_3$&&\\[20pt]
\includegraphics[width=0.16\textwidth]{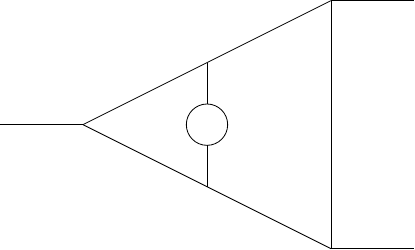}&$\Rightarrow$&&&\includegraphics[width=0.16\textwidth]{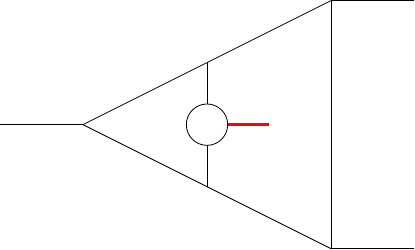}&$B_4$&&\\[20pt]
\includegraphics[width=0.16\textwidth]{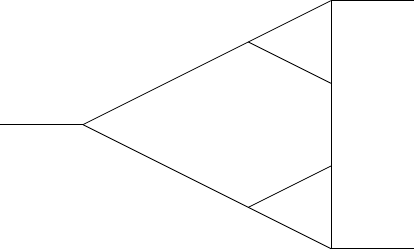}&$\Rightarrow$&&&\includegraphics[width=0.16\textwidth]{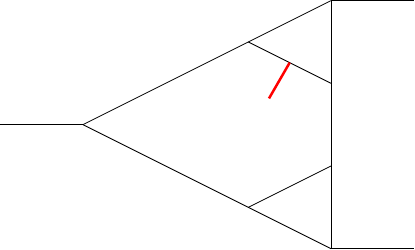}&$B_7^*$&&
\end{tabular}}

\label{Tab1:gluonicwebs}
\captionof{figure}{All non-planar top sectors for the leading-colour $H\to3$ partons amplitude are constructed by attaching an off-shell current (red) to three-point three-loop planar gluonic webs with 9 lines. Resulting planar topologies are omitted and reducible top sectors are indicated with an asterisk. The families $B_1, B_3, B_4, F_1, G_1$ and $I_2$ appear also in a three-point form factor of $\Tr(\phi^2)$ in planar $\mathcal{N}=4$ sYM.
}
\label{fig: gluonic webs}
\end{center}
\end{table}

The equivalent quantity in QCD arises in an effective field theory with $N_f$ light quarks and a very large 
top quark mass~\cite{Kniehl:1995tn,Chetyrkin:1997un} described by the Lagrangian
\begin{align}
    \mathcal{L}_{\mathrm{HEFT}} = -\frac{\lambda}{4}HG_a^{\mu\nu}G_{a,\mu\nu}\,,
    \label{eq:Heft}
\end{align}
where $G_a^{\mu\nu}$ is the gluonic field-strength tensor and $H$ the Higgs field. 
Consequently, the Higgs couples directly to gluons and all internal propagators are massless.
In the limit of large $N \sim N_f$, the graphs for both quantities are therefore produced by attaching the external current to a planar web of gluons (and quarks in QCD) with three external legs in all possible ways. 
On top of the known planar configurations, this allows for 9 different non-planar graphs with the maximum number of 10 internal lines, which we organise into integral families $B_1, B_3, B_4, B_5, B_6, B_7, F_1, G_1, I_2$, see fig.~\ref{fig: gluonic webs}. 
Note that the master integrals for family $B_1$ have already been computed in~\cite{Henn:2023vbd}.

The integrands can be written in terms of a combination of scalar integrals of 
the form
\begin{align}
  \mathrm{INT}(a_1,...,a_{15}) =  (-q^2)^{-3\e}e^{3\gamma_E \e} \int\left(\prod_{l=1}^3\frac{d^dk_l}{i\pi^{d/2}}\right)\prod_{i=1}^{10}P_i^{-a_i}\prod_{i=11}^{15}N_i^{a_i}
    \label{eq:feynman integral definition}
\end{align}
with $k_l$ the three loop momenta, $P_i$ the internal propagators which define the \textit{integral family} and $N_i$ the auxilliary numerators.
The lists of propagators and auxilliary numerators in the 9 non-planar families considered here are given in appendix~\ref{app: A}.
A \textit{sector} is then defined by a characteristic subset of the propagators $P_i$ and we refer to any sector with $t=10$ internal propagators as a \textit{top sector}.

All scalar Feynman integrals in the families in fig.~\ref{tab: integral families}
can be reduced  to a small set of master integrals using IBP identities~\cite{Chetyrkin:1981qh,Laporta:2000dsw}.
We evaluate the master integrals with the method of differential equations~\cite{Kotikov:1991pm,Remiddi:1997ny,Gehrmann:1999as}. 
In~\cite{Henn:2013pwa}, it was conjectured that Feynman integrals that are pure and UT satisfy a system of differential equations in the \textit{canonical} form
\begin{align}
d \vec{J}(\vec{x};\e) &= \e \sum_{j} \, d \log{(l_j)}  M_j  \, \vec{J}( \vec{x};\e)\nonumber \\
&= \e \,  \mathrm{d}\tilde{M}(\vec{x})  \, \vec{J}( \vec{x};\e),
\label{eqs:canonical_eq}
\end{align}
where $l_j = l_j(\vec{x})$ are the alphabet letters corresponding to the differential forms $\omega_i$ of sec.~\ref{sec: functions} and $M_j$ are matrices of rational numbers. 

The canonical form \eqref{eqs:canonical_eq} has a number of positive attributes. 
The analytic structure of the integrals becomes transparent and determines the functional space of its solutions, 
whereas the $\e$-factorisation property allows for a straightforward solution of 
the system of differential equations at each order in $\e$ in terms of iterated integrals defined in \eqref{eqs:Chen_int}. 
A~formal solution of the DEs in~\eqref{eqs:canonical_eq} is given by the path-ordered exponential
\begin{align}
\vec{J}(\vec{x},\e) = \mathbb{P} e^{\e \int_{\gamma} \mathrm{d} \tilde{M}}\,\vec{J}_0(\e)\,,
\label{eqs:pathexp}
\end{align}
where the path $\gamma$ connects the base point previously indicated with $\vec{1}_0$ to an arbitrary kinematic point $\vec{x}$. 
Expanding~\eqref{eqs:pathexp} in powers of $\e$, the solution is written
order by order in $\e$ as a combination of iterated integrals over the $d\log$ kernels appearing in the matrix $\mathrm{d}\tilde{M}$, see~\eqref{eqs:Chen_int}. 
In the remainder of this section, we will describe the construction of canonical differential equations for all relevant master integrals and discuss in detail their analytic structure. 

\subsection{Finding canonical candidates}
\begin{table}[]
\centering
\resizebox{0.495\textwidth}{!}{
\begin{tabular}[t]{c|c|c}
\textbf{irreducible} family & \# masters & new letters\\ \hline \hline
\addstackgap[4pt]{\makecell{$B_1$ \\  \includegraphics[width=0.23\textwidth]{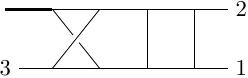}}}    & \makecell{150 \\ (4 in top)}  & \makecell{$l_{12}$ at  $\mathcal{O}(\epsilon^4)$ \\ $l_{13}$ at  $\mathcal{O}(\epsilon^4)$} \\ \cline{2-3}
\addstackgap[4pt]{\makecell{$B_3$ \\  \includegraphics[width=0.23\textwidth]{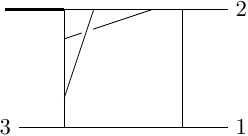}}}     & \makecell{90 \\ (1 in top)} & $l_{7, 8}$ at  $\mathcal{O}(\epsilon^5)$ \\ \cline{2-3}
\addstackgap[4pt]{\makecell{$F_1$ \\  \includegraphics[width=0.23\textwidth]{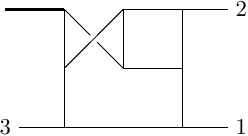}}}     & \makecell{214 \\ (5 in top)}& \makecell{$l_{7, 8}$ at  $\mathcal{O}(\epsilon^4)$ \\ $l_{10}$ at  $\mathcal{O}(\epsilon^5)$}  \\ \cline{2-3}
\addstackgap[4pt]{\makecell{$G_1$ \\  \includegraphics[width=0.23\textwidth]{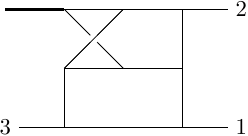}}}     & \makecell{254 \\ (6 in top)}& \makecell{$l_{7, 8}$ at  $\mathcal{O}(\epsilon^4)$\\ $l_{9}$ at  $\mathcal{O}(\epsilon^5)$\\ $l_{11}$ at  $\mathcal{O}(\epsilon^4)$}     \\ \cline{2-3}
\addstackgap[4pt]{\makecell{$I_2$ \\  \includegraphics[width=0.23\textwidth]{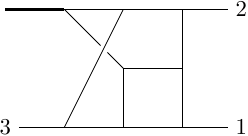}}}     & \makecell{305 \\ (8 in top)}& \makecell{$l_{7, 8}$ at  $\mathcal{O}(\epsilon^5)$\\ $l_{9}$ at  $\mathcal{O}(\epsilon^5)$\\ $l_{10}$ at  $\mathcal{O}(\epsilon^5)$\\ $l_{11}$ at  $\mathcal{O}(\epsilon^4)$\\ $l_{12}$ at  $\mathcal{O}(\epsilon^4)$\\}    \\ 
\end{tabular}
}
\resizebox{0.495\textwidth}{!}{
\begin{tabular}[t]{c|c|c}
\textbf{reducible} family & \# masters & new letters\\ \hline \hline
\addstackgap[4pt]{\makecell{$B_4$ \\  \includegraphics[width=0.23\textwidth]{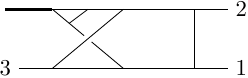}}}     & \makecell{143 \\ 
(0 in top)}& $l_{7, 8}$ at  $\mathcal{O}(\epsilon^4)$\\ \cline{2-3}
\addstackgap[4pt]{\makecell{$B_5$ \\  \includegraphics[width=0.23\textwidth]{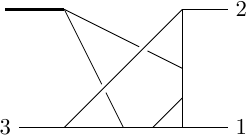}}}     & \makecell{70 \\ (0 in top)} & \xmark     \\ \cline{2-3}
\addstackgap[4pt]{\makecell{$B_6$ \\  \includegraphics[width=0.23\textwidth]{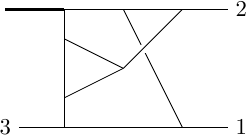}}}     & \makecell{150 \\ (0 in top)}& $l_{9}$ at   $\mathcal{O}(\epsilon^5)$\\ \cline{2-3}
\addstackgap[4pt]{\makecell{$B_7$ \\  \includegraphics[width=0.23\textwidth]{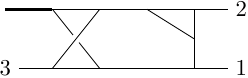}}}     & \makecell{89 \\ (0 in top)} & $l_{11}$ at   $\mathcal{O}(\epsilon^4)$ \\ 
\end{tabular}}

\caption{An overview of integral families computed in this paper, indicating the number of master integrals and the $\epsilon$ order at which new three-loop letters (see table~\ref{tab:alphabet}) appear for the first time. Master integrals are normalised to start at $\mathcal{O}(\epsilon^0)$ with a rational constant.}
\label{tab: integral families}
\end{table}
While many techniques have been developed in the past decade, 
finding a canonical basis remains only a partially automated task which requires considerable manual input.
As for most cutting-edge calculations, no single technique was enough to address all relevant graphs and we used a combination of various tools.
We first obtained an initial basis of master integrals of the form~\eqref{eq:feynman integral definition} 
for every irreducible sector appearing in the integrand using \texttt{Kira2}~\cite{Klappert:2020nbg} and \texttt{FIRE6}~\cite{Smirnov:2023yhb}, which implement and automate the Laporta algorithm~\cite{Laporta:2000dsw}. 
After mapping all possible subtopologies to the known planar canonical bases of~\cite{DiVita:2014pza,Canko:2021xmn} with \texttt{Reduze}~\cite{vonManteuffel:2012np}, we populated the remaining sectors of the families in table~\ref{tab: integral families} using the following methods.

The first approach is to look for integrals that are by design pure and UT, which is by conjecture satisfied for integrals with constant \textit{leading singularity}~\cite{Henn:2013pwa}.
Starting from an individual Feynman integral, we introduce a numerator ansatz which effectively produces a 
linear combination of integrals belonging to the given sector and its subsectors.
The ansatz is constrained by requiring that the canonical candidate is UV finite, 
that its integrand is free of double poles in any of the integration variables, and that the leading singularities 
of its maximal cut
(i.e. the maximally iterated residues of the integrand where all propagators have been put on shell) 
are constant~\cite{Henn:2020lye}.
In complicated cases, this analysis might not provide a fully canonical basis, but it always supplies candidates where at least the homogeneous part of their differential equations is in canonical form.
This is due to the fact that the maximal cut of a Feynman integral is a solution of the homogeneous part of the differential equation it satisfies~\cite{Primo:2016ebd}. 
As can be seen from \eqref{eqs:canonical_eq}, in the limit $\e \to 0$, the canonical differential equation becomes trivial, implying that the homogeneous solution for $\e = 0$ is a constant. 

The success of the above analysis is often strongly dependent on the adopted parametrisation of loop momenta. 
One choice is the Baikov representation~\cite{Baikov:1996iu, Baikov:1996rk}, implemented for example in the package \texttt{Baikov.m}~\cite{Frellesvig:2017aai}, where the propagators of the integral take the role of the integration variables and setting them to zero produces the residues.
The loop-by-loop Baikov representation~\cite{Frellesvig:2017aai} can additionally 
be used to reduce the number of integration variables. 
In our case, studying the maximal cut using either the loop-by-loop Baikov representation or in momentum space~\cite{Flieger:2022xyq} provided candidates for the hardest top sectors. 
For many top sectors, we had to extend the leading-singularity analysis beyond the maximal cut to fix the non-homogeneous part of the differential equation (see appendix B of~\cite{Baranowski:2022vcn} for an explicit example).
Another popular choice of parametrisation relies on spinor-helicity variables in $d=4$ dimensions and is implemented in the package \texttt{DlogBasis}~\cite{Henn:2020lye}. 
This purely four-dimensional representation is intrinsically different from the 
Baikov representation and it was effective in supplying canonical candidates for most sectors with up to nine propagators. 

In the most complicated cases, we could only derive preliminary candidates $\vec{J}_0$ such that the subsectors $\vec{S}$ and the homogeneous part of the top sector $\vec{T}_0$ were in canonical form. 
For such a preliminary basis of integrals
\begin{equation}
    \vec{J}_0 = \big(\vec{T}_0,\, \vec{S} \big)\,,
\end{equation}
the differential equations take the form
\begin{equation}
 d\vec{J}_0=\left(\begin{array}{c|cccccc}
    \e\, B &  &\quad \tilde{B}(\e)& & & &  \\ \hline
      & \multicolumn{5}{c}{\multirow{3}{*}{\raisebox{0mm}{\scalebox{1}{$\e\, C$}}}} \\
    \raisebox{2pt}{0} & & &\\
     & & & 
  \end{array}\right)\cdot
  \left(\begin{array}{c}
    \vec{T}_0 \\ \hline
    \\
   \vec{S} \\
    \\
  \end{array}\right).
\end{equation}
If $n$ is the number of top sector basis elements and $m$ is the number of subsector basis elements, 
$\tilde{B}(\e)$ is a $n\times m$ matrix which we could always choose as at most linear in $\e$, 
$\tilde{B}(\e) = \tilde{B}_0 +\e \tilde{B}_1 $. 
In this case, we followed the procedure described in~\cite{Gehrmann:2014bfa,Argeri:2014qva}, solved the system for $\e =0$ and derived a transformation matrix that would realise the $\e$-factorised form.

Finally, we computed the DEs for the candidate basis with a combination of IBP tools~\cite{vonManteuffel:2012np,Lee:2013mka,Klappert:2020nbg,Peraro:2019svx}. 
In order to avoid expensive reductions of individual integrals, we opted for direct reconstruction of the differential equations following~\cite{Henn:2023vbd,Syrrakos:2023mor}.
The final canonical bases for all relevant non-planar families are described in table~\ref{tab: integral families} and provided in the supplementary material.

\iffalse
\renewcommand{\arraystretch}{1.05}
\begin{table}[]
\centering
\begin{tabular}{l|l|l|l}
\multicolumn{1}{c|}{two-loop}  & \multicolumn{1}{c|}{square roots} & \multicolumn{1}{c|}{quadratic}  & \multicolumn{1}{c}{kinematic crossings} \\ \hline
\multicolumn{1}{l|}{ $l_1 = x$  } & \multicolumn{1}{l|}{\addstackgap[4pt]{$l_7 = \frac{(1-x-y) x - \sqrt{-x (1-x-y) y}}{(1-x-y) x + \sqrt{ -x (1-x-y) y}}$}}  & $l_9 = 1-y\left(1+x\right)$ & $l_{14} = \left(x+y\right)^2-y$ \\ 
\multicolumn{1}{l|}{ $l_2 = y$ } & \multicolumn{1}{l|}{\addstackgap[4pt]{$l_8 = \frac{x y - \sqrt{- x (1-x-y) y}}{x y + \sqrt{-x (1-x-y) y}}$}}  & $l_{10} = 1-x\left(1+y\right)$ & $l_{15} = \left(x+y\right)^2-x$ \\ 
\multicolumn{1}{l|}{$l_3 = 1-x-y$} & \multicolumn{1}{l|}{} & $l_{11} = y^{2}+x-y$ & $l_{16} = \left(y-1\right)^2-x$ \\  
\multicolumn{1}{l|}{$l_4 = 1-x$} & \multicolumn{1}{l|}{} & $l_{12} = x^{2}+y-x$ & $l_{17} = y^2+x y+x$ \\ 
\multicolumn{1}{l|}{$l_5 = 1-y$} & \multicolumn{1}{l|}{} & $l_{13} = -y+\left(1-x\right)^2$ & $l_{18} = x^2+x y+y$ \\ 
\multicolumn{1}{l|}{$l_6 = x+y$} &  & & $l_{19} = \left(y-1\right)^2+x y$ \\ 
 &  & & $l_{20} = \left(x-1\right)^2+x y$ \\ 
\end{tabular}
\label{tab:alphabet_old}
\caption{Symbol letters appearing in the differential equations for the topologies considered in this paper, and their kinematic crossings.}
\end{table}
\fi

\begin{figure}
\centering
\begin{subfigure}{\textwidth}
\centering
\begin{tabular}{c|ll}
permutation & \multicolumn{2}{c}{linear letters}\\ \hline \hline
\textbf{()} & $\bm{l_1 = x}$& $\bm{l_4 = 1-x}$\\ \hline
(23)  & $l_2 = y$& $l_5 = 1-y$\\
(13)  & $l_3 = 1-x-y$  & $l_6 = x+y$\\
(12)  & $\color{gray}{l_1}$  & $\color{gray}{l_4}$ \\
(123) &  $\color{gray}{l_3}$ & $\color{gray}{l_6}$ \\
(132) & $\color{gray}{ l_2}$ & $\color{gray}{l_5}$
\end{tabular}
 \label{tab:alphabet2L}
\subcaption{Alphabet letters of the one- and two-loop four-point master integrals with one off-shell leg.}
\end{subfigure}
\par\bigskip
\begin{subfigure}{\textwidth}
\renewcommand{\arraystretch}{1.2}
\centering
\begin{tabular}{c|l|l|l}
permutation & \multicolumn{1}{c|}{parabolic let.}& \multicolumn{1}{c|}{hyperbolic let.}& \multicolumn{1}{c}{square roots}\\ \hline \hline
\textbf{()}  & $\bm{l_{11} = y^2-y+x}$& $\bm{l_{17} = y^2+xy+x}$& \multicolumn{1}{l}{\addstackgap[4pt]{$\bm{l_8 = \frac{x y - \sqrt{- x (1-x-y) y}}{x y + \sqrt{-x (1-x-y) y}}}$}} \\ \hline
(23)    & $l_{12} = x^2-x+y$ & $l_{18} = x^2+xy+y$& $\color{gray}{l_8}$\\
(13)   & $l_{16} = \left(1-y\right)^2-x$& $l_{10} = 1-x\left(1+y\right)$  & $\color{gray}{(l_7\cdot l_8)^{-1}}$\\
(12)    & $l_{14} = \left(x+y\right)^2-y$& $l_{19} = \left(y-1\right)^2+x y$ & \multicolumn{1}{l}{\addstackgap[4pt]{$l_7 = \frac{(1-x-y) x - \sqrt{-x (1-x-y) y}}{(1-x-y) x + \sqrt{ -x (1-x-y) y}}$}}\\
(123)   & $l_{13} = \left(x-1\right)^2-y$& $l_{9\phantom{0}} = 1-y\left(1+x\right)$ & $\color{gray}{l_7}$ \\
(132)   & $l_{15} = \left(x+y\right)^2-x$& $l_{20} =  \left(x-1\right)^2+x y$ & $\color{gray}{(l_7\cdot l_8)^{-1}}$
\end{tabular}
 \label{tab:alphabet3L}
\subcaption{The extension of the alphabet at three loops. }
\end{subfigure}
\captionof{table}{We denote the 6 permutations of the external legs $p_1, p_2, p_3$ as elements of the group $S_3$ with cycle notation. Only the letters $l_{1},\ldots,l_{13}$ appear in the uncrossed integral families defined in table~\ref{tab: integral families}. The crossing symmetry in $(x,y,z)$ between the letters of one type is not always manifest since we eliminate one variable, $z \to 1-x-y$.}
\label{tab:alphabet}
\end{figure}

\subsection{Three-loop alphabet}
\label{subsec: alphabet}
\begin{figure}[]
\centering
\includegraphics[width=0.39\linewidth]{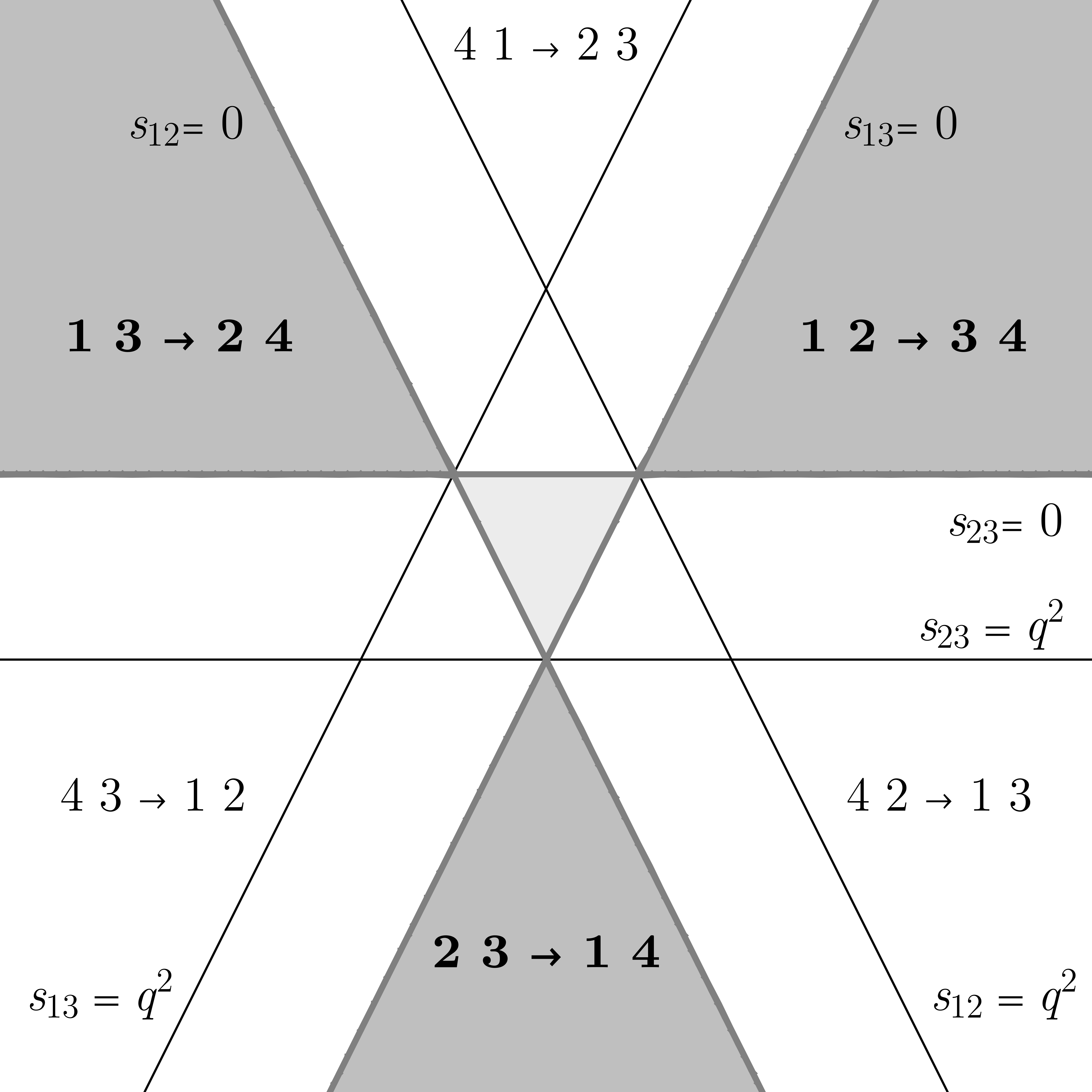}
\caption{The kinematic regions of four-point amplitudes with one off-shell leg. The central light grey triangle delimited by the lines $s_{12}=0, s_{13}=0, s_{23}=0$ is the decay region. The dark grey shaded regions are the $s_{12},s_{13},s_{23}$ channels of $2\to2$ scattering. In these regions, $q^2>0$. The remaining three white regions have deep-inelastic-scattering-like kinematics and $q^2<0$. The black lines also represent the curves where the letters $l_{1-6}$ vanish.}
\label{Fig: kinematicPlane}
\end{figure}
\begin{figure}[h]
\centering
\includegraphics[width=0.49\linewidth]{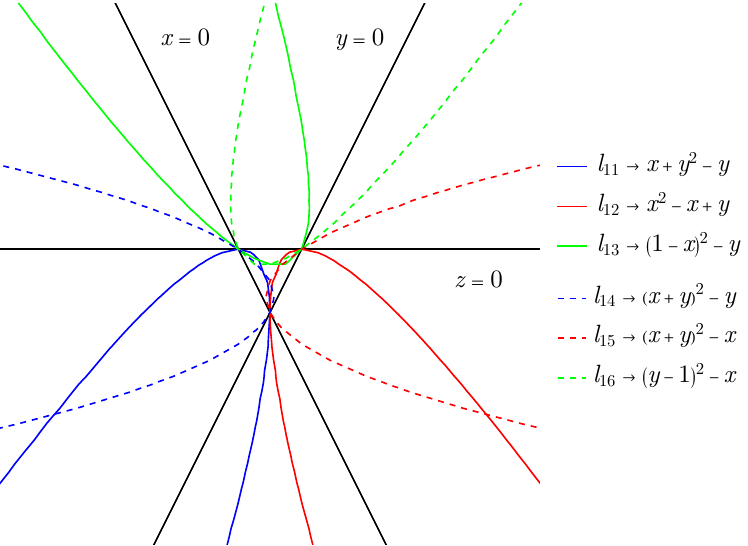}
\includegraphics[width=0.49\linewidth]{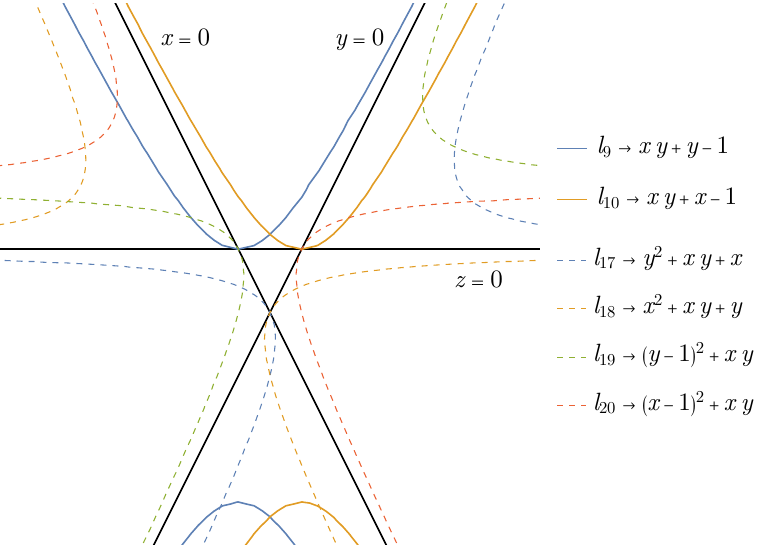}
\caption{Curves of vanishing letters which appear at 3 loops for the first time. We distinguish two groups: parabolic (left) and hyperbolic (right). Letters appearing in our families are in full line, dashed lines represent letters which appear only in kinematic crossings of the integral families. We do not display the square root letters, which only vanish in the DIS-like regime along the lines $y=1$, $x=1$ and $z=1$.}
\label{Fig: singular}
\end{figure}
The derivation of canonical DEs reveals the analytic structure of the ensuing iterated integrals.
Up to 2 loops, and in the planar 3-loop integrals, only the three kinematic crossings of the two forms $x$ and $1-x$ appeared.
In non-planar topologies at three loops, we observe for the first time three new categories of letters, see table~\ref{tab:alphabet}.
First of all, there are two types of quadratic letters, 
corresponding to the 6 kinematic crossings of the two polynomials $y^2-y+x$ and $y^2+xy+x$.
We dub these letters \textit{parabolic} and \textit{hyperbolic} after the curves in the $(x,y)$-plane that they define. 
When written in terms of a pair of the symmetric variables $x,y,z$, 
the latter are always linear in one of the variables. 

Secondly, we encounter two algebraic letters which correspond to two crossings 
of the same expression involving the square root $\sqrt{- s_{12}\,s_{13}\,s_{23}\,q^2}$, which in terms of $x,y,z$
read
\begin{align}
\left\{ \frac{x\,y - \sqrt{-x\,y\,z}}{x\,y + \sqrt{ -x\,y\,z}} ,\frac{ x\,z - \sqrt{-x\,y\,z}}{ x\,z + \sqrt{ -x\,y\,z}} \right\}\,.
\end{align}
Interestingly, the third crossing can be expressed in terms of the first two, 
\begin{align}
\frac{y\,z - \sqrt{-x\,y\,z}}{y\,z + \sqrt{ -x\,y\,z}} = \left[
\frac{x\,y - \sqrt{-x\,y\,z}}{x\,y + \sqrt{ -x\,y\,z}} \cdot \frac{ x\,z - \sqrt{-x\,y\,z}}{ x\,z + \sqrt{ -x\,y\,z}} \right]^{-1}\,.
\end{align}
It is possible to rationalise the root, but only at the price of transforming 
some of the other letters into quartic polynomials in the new variables. 
For this reason, we keep the square root as it is.

The kinematic plane is depicted in  fig.~\ref{Fig: kinematicPlane} in equilateral coordinates.
The (Euclidean) decay region corresponds to the central light grey triangle, while the scattering regions are indicated by darker grey areas. 
In fig.~\ref{Fig: singular}, we show the singularities corresponding to the new three-loop alphabet letters. 
Note that as $q^2 \rightarrow 0$, all letters degenerate onto HPLs, the function space of four-point all-massless scattering up to three loops.

Inspecting the solution written in terms of Chen iterated integrals obtained by expanding eq.~\eqref{eqs:pathexp} in powers of $\e$, we can determine the first order at which a given integration kernel appears in the solution.
With our choice of basis, the algebraic and quadratic letters appear for the first time at $\mathcal{O}(\e^4)$ or $\mathcal{O}(\e^5)$. 
We notice that the quadratic letters never mix with each other nor with the square root letters in any single iterated integral up to the last $\e$-order relevant for the finite part of the amplitude. 
A detailed description of the appearance of new letters in each family is given in table~\ref{tab: integral families}.
Moreover, in our problem, it is easy to obtain the \textit{symbol} solution from the iterated integral representation, since products of transcendental constants and lower-weight functions belong to the kernel of the symbol map.
For four-point integrals with one off-shell leg, the following conditions on the sequence of letters appearing in the symbol have been put forward in~\cite{Dixon:2020bbt}, and also in~\cite{Chicherin:2020umh}, inspired by a description using the B2/C2 cluster algebras.
Referring to the indices of the letters $l_i$ in table~\ref{tab:alphabet}, they can be checked on the level of the coefficient matrices to all orders in $\e$ as
\begin{align}
    M_i \cdot M_j = M_j \cdot M_i = 0\,,\qquad i,j\in\{4,5,6\}, i\neq j\,,
    \label{eq:old adjacency}
\end{align}
which at three loops are known to be violated even for the planar integrals~\cite{Henn:2023vbd} but they are satisfied for example in the ensuing leading colour $V$+jet amplitude~\cite{Gehrmann:2023zpz}. 
For the new non-planar integrals, we again observe terms which do not obey these conditions. 
Interestingly, among the iterated integrals which contain the new 3L letters, only those with hyperbolic or square root letters violate the conditions~\eqref{eq:old adjacency}.

In addition, seven new relations have been recently proposed based on associating part of the alphabet with the cluster algebra G2~\cite{Aliaj:2024zgp}:
\begin{align}
    M_i \cdot M_j &= 0\,,\qquad (i,j)=(5,12),(5,13), (6,12), (6,13), (12,13)\,
        \label{eq:new adjacency1}
\end{align}
and
\begin{align}
    M_{12} \cdot (M_1+M_2+M_6+M_{12})&=0\,,\\
    (M_1+M_3+M_5+M_{13}) \cdot M_{13}&=0\,.
    \label{eq:new adjacency2}
\end{align}
We find that these conditions and their kinematic crossings are satisfied for all the integral families, and therefore for any amplitude, even beyond the symbol level.

\begin{figure}
\hspace{2.4cm}\textbf{square roots}\qquad\qquad\quad\quad\qquad \textbf{hyperbolic} \quad\qquad\quad \textbf{parabolic}
\\[6pt]
\centering
\begin{subfigure}{.24\textwidth}
  \centering
  \includegraphics[width=.9\linewidth]{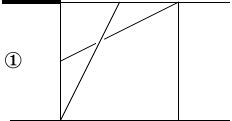}
\end{subfigure}
\begin{subfigure}{.24\textwidth}
  \centering
  \includegraphics[width=.9\linewidth]{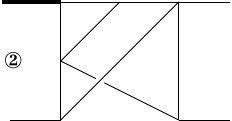}
\end{subfigure}
\begin{subfigure}{.24\textwidth}
  \centering
  \includegraphics[width=.9\linewidth]{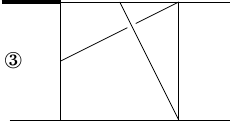}
\end{subfigure}
\begin{subfigure}{.24\textwidth}
  \centering
  \includegraphics[width=.9\linewidth]{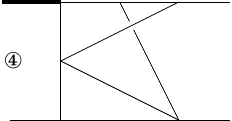}
\end{subfigure}
\caption{The four diagrams with 8 propagators which are the sources of new letters first appearing at 3 loops: \ding{172} and \ding{173}~are the lowest topologies containing a square root $l_{7,8}$, \ding{174}~the hyperbolic letter $l_{18}$, \ding{175}~the parabolic letter $l_{12}$, as defined in table~\ref{tab:alphabet}. Note that each diagram appears in all its six kinematic crossings.}
\label{fig: sources}
\end{figure}

The occurrence of quadratic and square root letters can be associated with just four diagrams with 8 internal lines (see fig.~\ref{fig: sources}). 
These diagrams have the given letter in the homogeneous part of their DE, i.e. $\sqrt{-xyz}$ is part of their leading singularity. 
Some of their supersectors develop the new letters from the next $\e$ order due to a coupling in the differential equations.

A systematic study of analytic continuation from the decay region to the scattering regions is beyond the scope of the present work. 
Nonetheless, the knowledge of the full alphabet allows us to make some observations.
The analytic continuation of polylogarithmic functions with an alphabet given by the first six letters in table~\ref{tab:alphabet} is well-understood~\cite{Gehrmann:2002zr}. 
Namely, different changes of variables to access all scattering regions have been devised such that the alphabet transforms back into itself. 
For example, the change of variables from the decay region to the region describing the scattering process $p_1 + p_2 \rightarrow p_3 + p_4$ is given by
\begin{align}
 x = \frac{1}{v}\,, \qquad y = -\frac{u}{v}
 \label{eqs: an_cont}
\end{align}
with $0 \leq v \leq 1$, $0 \leq u \leq 1-v $. 
All other scattering regions are subsequently accessible just by kinematic crossing.

Interestingly, we find that the quadratic sector of the new alphabet is still mapped onto itself under transformations of the form of eq.~\eqref{eqs: an_cont}.
Each scattering region is traversed by singularities associated with two parabolic and four hyperbolic letters, see fig.~\ref{Fig: singular}. 
Transformations of the form~\eqref{eqs: an_cont} map the two parabolic letters entering a given scattering region into two parabolic letters in the new variables, while the other four parabolic letters are mapped to hyperbolic letters.
On the other hand, the four hyperbolic letters entering the scattering region are mapped to four parabolic ones and the remaining two hyperbolic ones are mapped into two hyperbolic letters. 
Explicitly, consider the transformation in eq.~\eqref{eqs: an_cont} followed by the renaming $v \rightarrow y$, $u \rightarrow x$. 
In this case, the mapping acts on the parabolic letters as
\begin{align}
(l_{11}, l_{12}, l_{13}, l_{14}, l_{15}, l_{16}) & \rightarrow ( \underbrace{l_9, l_{18}, l_{19}, l_{20}}_{\text{hyperbolic}} , \underbrace{l_{13}, l_{14}}_{\text{parabolic}})\,,
\end{align}
and on the hyperbolic letters as
\begin{align}
(l_{10}, l_{17}, l_{19},l_{20},l_{9},l_{18}) & \rightarrow (\underbrace{l_{12},l_{13},l_{15},l_{16}}_{\text{parabolic}},\underbrace{l_{17},l_{10}}_{\text{hyperbolic}})\,.
\end{align}
The square root letters are not closed under this transformation, but as we will see in subsection~\ref{subsec: Hjet results}, they are not relevant for the leading-colour $H$+jet amplitudes.
\section{Boundary conditions}
\label{sec: boundaries}
The formal solution of the differential equation system introduced in section~\ref{sec: functions} requires fixing values of the integrals at the base point in the path-ordered exponential solution of eq.~\eqref{eqs:pathexp}. 
We determine all boundary constants analytically by enforcing the correct behaviour of the integrals at a number of kinematic surfaces. Typically, physical requirements such as regularity at pseudo-thresholds or asymptotic behaviour close to physical thresholds furnish enough conditions to completely determine all master integrals in terms of a few simple one-scale integrals that have to be computed with other methods.
Unfortunately, these conditions are given at disparate phase-space points, often lying outside of a convenient Euclidean region where all integrals are real and single-valued. 
Full analytic control on the iterated integrals is therefore required to access these points, including the ability to take limits and analytically continue the integrals to arbitrary kinematic points.

As discussed in section~\ref{subsec: alphabet}, up to two loops and in the 
three-loop planar case, all master integrals with three on-shell and one off-shell leg can be expressed to all orders in $\e$ as iterated integrals involving only the first $6$ letters of the alphabet defined in table~\ref{tab:alphabet}, $l_{1-6}$.
The linearity of these letters allows one to express all the iterated
integrals in terms of MPLs~\cite{Goncharov:1998kja,Remiddi:1999ew} at once, rendering their manipulation straightforward with standard methods, as implemented for example in \texttt{PolyLogTools}~\cite{Duhr:2019tlz}. 
In practice, one can then fix the boundary values at every order in $\e$ by explicitly taking limits of these functions to the relevant regular or singular regions. 

In contrast, the appearance of non-linear and algebraic letters $l_{7-20}$ 
in the non-planar three-loop integrals significantly complicates the analytic treatment of the resulting functions. 
Since our goal at this point is to harness the grading introduced in section~\ref{sec: functions} and identify only the functions relevant to the amplitude, we prefer to fix boundary constants prior to devising a full functional representation.
To this end, it is sufficient to solve a simpler one-dimensional problem. 
Instead of fixing all boundary conditions for the full solution, one can in fact solve the differential equations along a set of one-dimensional paths which are specifically chosen to enable transporting the information given at the various thresholds and pseudo-thresholds to the common base point $\vec{1}_0$. 
Moreover, with an appropriate parametrisation, it is often possible to linearise the alphabet of the resulting differential equations, which in turn renders their solution in terms of MPLs straightforward.

Specifically, we start with identifying the kinematic slices associated with a vanishing alphabet letter, corresponding either to a physical threshold or a pseudo-threshold in the DEs. 
Next, we perform a change of variables to specify some path $\gamma$ asymptotically close to each slice, ensuring that the paths are connected to each other and to 
the base point.
On each slice, the DEs contain fewer variables and the alphabet simplifies, which allows us to solve the equations in terms of a reduced number of variables and impose the boundary conditions accessible in that region. 

Enforcing the compatibility of the partial solutions on the intersections of the paths, one can finally transport all the boundary information to the base point, where the value of all integrals is thereby fixed to a number.
Importantly, for this strategy to succeed, we have to make sure that
the linearity of the alphabet is preserved at each intersection, such that all boundary constants can be effectively transported to the base point in terms of 
MPLs only. 
This point is particularly delicate, especially when considering paths
that meet at the intersection of more than two singular slices. 
In these cases, one cannot perform the matching simply by taking the limit of each path, since this procedure is typically ill-defined.
Instead, a well-defined path must be followed to avoid the ambiguities originating
from crossing multiple singular surfaces at once.\footnote{This problem is well-understood mathematically and our solution is equivalent to performing a suitable
\emph{blow-up} at the point where the singular slices intercept.}
If such a set of singular surfaces and connecting paths where the alphabet linearises can be identified, all the boundary conditions can effectively be transported to a common base point, bypassing the need for a general solution of the differential equations in the full kinematic plane.

\begin{figure}[h]
\noindent\begin{minipage}{\linewidth}
\centering
\includegraphics[width=0.7\textwidth]{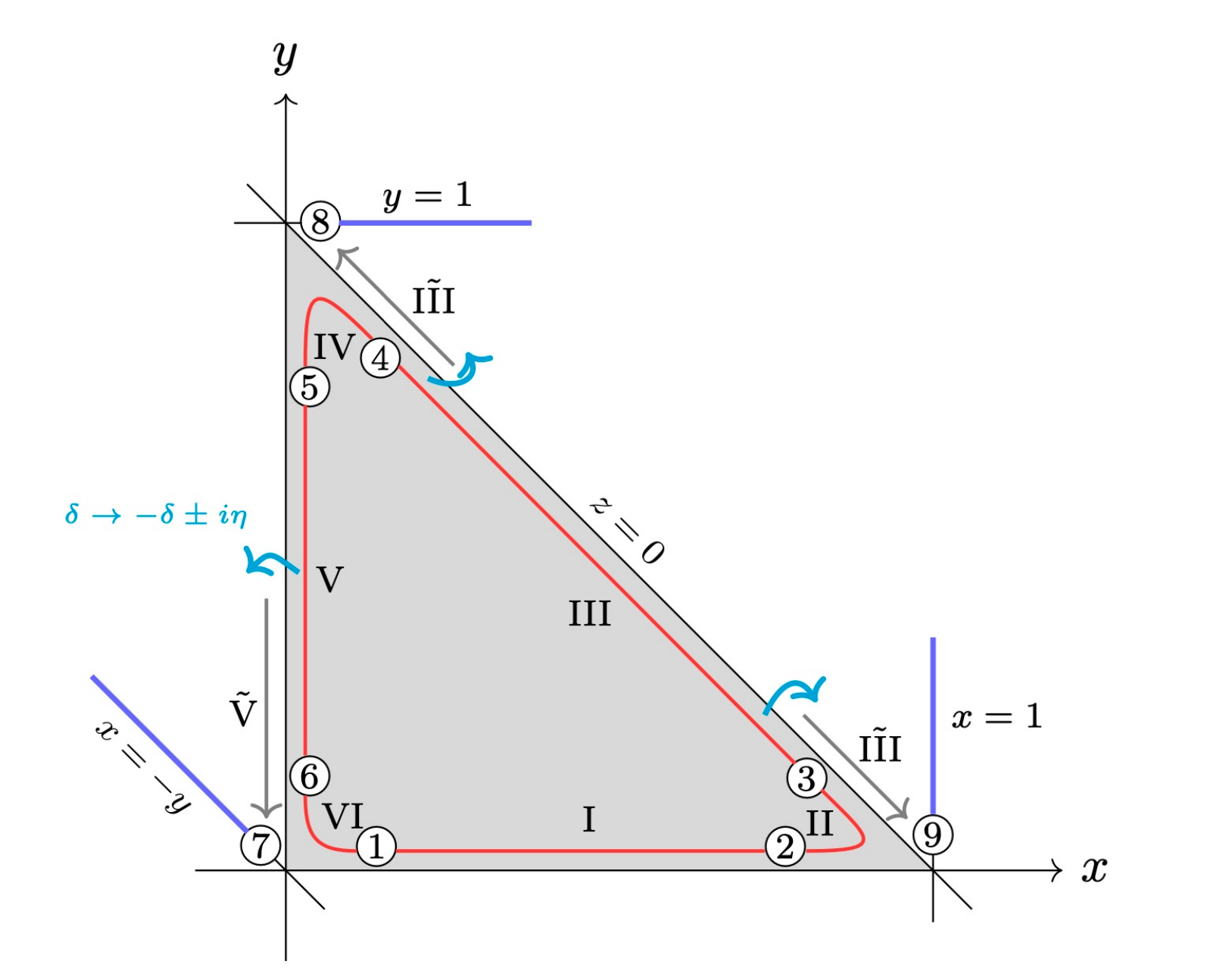}
\renewcommand{\arraystretch}{1.05}
\centering
\begin{tabular}{llll}
\multicolumn{1}{l}{segment} & start-/end-point & \multicolumn{1}{l}{$x(t,\delta)$ } & \multicolumn{1}{l}{$y(t,\delta)$ }\\ \hline
\multicolumn{1}{l}{ $\quad$  I} & $P_1 \rightarrow P_2$& \multicolumn{1}{l}{$t$} & \multicolumn{1}{l}{$\delta$}   \\ \hline
\multicolumn{1}{l}{ $\quad$  II} & $P_2 \rightarrow P_3$ & \multicolumn{1}{l}{$1-\delta \left[(1-t)^2 + t^2\right]$} & \multicolumn{1}{l}{$\delta t^2$}  \\ \hline
\multicolumn{1}{l}{ $\quad$  III} & $P_3 \rightarrow P_4$& \multicolumn{1}{l}{$1-t-\delta$} & \multicolumn{1}{l}{$t$} \\ \hline
\multicolumn{1}{l}{ $\quad$  IV} & $P_4 \rightarrow P_5$& \multicolumn{1}{l}{$\delta (1-t)^2$} & \multicolumn{1}{l}{$1-\delta \left[(1-t)^2 + t^2)\right]$}  \\ \hline
\multicolumn{1}{l}{ $\quad$  V} & $P_5 \rightarrow P_6$& \multicolumn{1}{l}{$\delta$} & \multicolumn{1}{l}{$1-t$}  \\ \hline
\multicolumn{1}{l}{ $\quad$  VI}& $P_6 \rightarrow P_1$ & \multicolumn{1}{l}{$\delta t^2$} & \multicolumn{1}{l}{$\delta (1-t)^2$}  \\ \hline
\end{tabular}
\captionof{figure}{Single-variable kinematic slices used in the determination of boundary constants. The Euclidean region is the triangle enclosed by black lines. The red segments I,II,III IV,V,VI are infinitesimally close to the branch lines. The blue segments correspond to regular slices where an infinitesimal displacement is not needed. Turquoise arrows are used to indicate analytic continuation outside of the Euclidean region to connect the red and blue segments. The table summarises the parametrisation of each segment in terms $(\delta, t)$ with values  $\delta\rightarrow 0^+$ and $  t\in(0,1)$. The extremal points of the segments $P_1, \dots, P_6$ are drawn in the figure in circles. Points $P_7, P_8, P_9$ enable matching between the inner triangle and the regular limits $x=-y, y=1, x=1$.}
\label{fig:segments boundary}
\end{minipage}
\end{figure}

\subsection{Application to four-point integrals with one off-shell leg}
In this subsection, we describe the application of this method to the calculation of four-point integrals with one off-shell leg. 
Let us consider the kinematic plane depicted in fig.~\ref{fig:segments boundary}, spanned by the two dimensionless variables $x$ and $y$ defined in eq.~\eqref{eqs:kinematic}.
We find that all boundary conditions can be fixed in terms of trivial one-scale two-point integrals by considering the behaviour of the master integrals close to the three singular surfaces $x = 0$, $y=0$, $z=0$, corresponding to the slices I, III, V and by imposing regularity at the three pseudo-thresholds $x=1$, $y=1$, $z=1$ (or $x=-y$), the blue segments in fig.~\ref{fig:segments boundary}. 
The three singular surfaces are connected with three additional paths, denoted as II, IV and VI in the same figure.

Each of these $9$ paths is parametrised by two variables $t$, $\delta$ as 
\begin{align}
\gamma(t,\delta) = \left( x(t,\delta), y(t,\delta)\right)\,.
\end{align}
In our notation, $0<t<1$ is the coordinate on the curve, while $\delta\to0^+$ parametrises the distance of the paths from the corresponding singular surface and acts therefore as a regulator. 
Our choice of parametrisation on each segment is indicated in the table below fig.~\ref{fig:segments boundary}.
With this choice, the alphabet on each segment reduces to a subset of
\begin{align}
    \vec{\alpha}_t = \Big\lbrace t, t-\frac{1}{2}, t-1, t-\frac{e^{i \pi/4}}{\sqrt{2}} ,t-\frac{e^{-i \pi/4}}{\sqrt{2}} , t \pm e^{i \pi/3}, t \pm e^{2 i \pi/3} \Big\rbrace\,,
    \label{eqs:regularised_alphabet}
\end{align}
which are all linear in $t$. 
Hence, a solution in terms of MPLs along these paths is straightforward.

Together the paths give access to all relevant limits and connect them to the base point $\vec{1}_0 = (0,0)$.
We considered the following conditions in fixing the boundary values of all master integrals:
\begin{itemize}
    \item We impose regularity at the three pseudo-thresholds $x \rightarrow 1$, $y \rightarrow 1$, $
    y\rightarrow -x$, corresponding to the letters $l_4$, $l_5$ or $l_6$ going to zero. 
    \item We obtain 
    more conditions by restricting the appearance of spurious eigenvalues
    in the limits $x,y,z \to 0$. In these limits, the differential equation~\eqref{eqs:canonical_eq} takes the simple form
     \begin{align}
d \vec{J}(\delta \sim 0;\e) = \e M_{\delta}\, d\log(\delta) \vec{J}(\delta \sim 0;\e)\,,
 \label{eqqs:asyeq}
\end{align}
    where $\delta$ represents the kinematic variable which approaches zero and $M_{\delta}$ is a numerical matrix. The solution of \eqref{eqqs:asyeq} is given by  $\delta^{\e M_{\delta}}\vec{J}_0(\e)$, with $\vec{J}_0$ a boundary vector. The matrix exponential contains terms
    of the form $\delta^{a \e}$ with the number $a$ positive or negative,
    but only the subset with $a<0$ is allowed. 
    This requirement can be related to the UV-finiteness of the canonical master integrals~\cite{Henn:2020lye}.
    \item These conditions relate all canonical integrals to simple single-scale integrals, such as the three-loop sunrise. 
    We computed closed form expressions for these integrals by direct integration using Feynman parameters.
\end{itemize}

Finally, we discuss the matching of boundary conditions at the intersection of segments. 
Referring to fig.~\ref{fig:segments boundary},
segments I to VI are connected and lie inside the Euclidean triangle, whereas the regular limits $x=-y, y=1, x=1$ lie outside of it. 
To link them, we define auxiliary paths, labelled in fig.~\ref{fig:segments boundary} as $\mathrm{\tilde{V}}$ and $\mathrm{\tilde{III}}$, which are parallel to V and III but infinitesimally outside of the Euclidean triangle. 
For example, solutions to the DEs on V and $\mathrm{\tilde{V}}$ are related by the analytic continuation $\delta \rightarrow - \delta \pm i \eta$, where the choice of the imaginary part $\eta$ is arbitrary.
Point $P_7$, infinitesimally close to the origin, can now be approached from segment $\tilde{V}$ and from the regular solution on surface $x=-y$. 
In an analogous way, we reach points $P_8$ and $P_9$. 
This choice of path to reach the limits outside of the Euclidean triangle is not unique.
As an alternative, one might for example perform analytic continuation from segment I to an auxiliary path $\mathrm{\tilde{I}}$ and from there again to the regular slices $x=-y$ and $x=1$.
Lastly, an ambiguity may arise in the definition of a regularised 
limit. 
In particular, matching of segments is often performed in the neighborhood of a singular point
where logarithmic divergences need to be integrated out. 
In Appendix~\ref{app: B}, we explain the solution to this technical issue with the help of an example. 

\subsection{Numerical checks}
We conducted a numerical check on the analytically determined boundary constants by comparing some of the canonical integrals against a numerical calculation performed with \texttt{AmFlow}.
The~kinematic point used in all evaluations was the symmetric point in the Euclidean region $(x_p, y_p)= (1/3,1/3)$, where all integrals are real-valued.
Our integral solutions were evaluated using a (generalised) power-series solution of the differential equation systems.
The expansion is centred around the point $(0,0)$ and written as a function of a single variable $r = x = y$. 
In order for the power series ansatz to capture the logarithmic behaviour of those integrals which diverge as $r \rightarrow 0$, we include at every weight $w$ all possible powers of the logarithm of $r$,
\begin{align}
\vec{J}_w(r) = \sum_{m,n} \vec{c}_{m,n}^{\,\,w}\, r^n (\log r )^m\,,
\label{eqs:gen_series}
\end{align}
where $0 \leq n \leq  n_{\text{max}}$ for a desired maximum power $n_{\text{max}}$ and $ 0 \leq m \leq w$. 
The coefficients $\vec{c}^{\,\,w}_{m,n}$ are fully determined by the boundary constants at $(0,0)$ and by requiring the ansatz~\eqref{eqs:gen_series} to satisfy the differential equation system. 

As described in subsection \ref{subsec: alphabet}, the letters appearing up to $\mathcal{O}(\e^3)$ are the same as at two loops and it is straightforward to obtain a solution in terms of MPLs.
Up to this order, the numerical evaluation is therefore very efficient and the agreement with \texttt{AmFlow} is up to 30 digits. 
The numerical values obtained using the generalised expansion at the symmetric point $(x_p, y_p)$ for the first integral in the canonical basis for each family are\newline
\resizebox{\textwidth}{!}{
\begin{minipage}{\textwidth}
\begin{alignat*}{16}
&\vec{J}_{\mathrm{B_1}}(1) &&=+0.277777 &&+0.915510 &&\e &&-1.911230 &&\e^2&&-27.71155 &&\e^3&&-126.0899 &&\e^4 &&-328.2572 &&\e^5&&-250.5231 &&\e^6, \\
&\vec{J}_{\mathrm{B_3}}(1) &&= && && &&-0.038299 &&\e^2&&+0.430934 &&\e^3&&-0.675799 &&\e^4 &&-12.48278 &&\e^5&&+4.351596 &&\e^6, \\
&\vec{J}_{\mathrm{B_5}}(1) &&= && && &&-0.274155 &&\e^2&&-1.304258 &&\e^3&&-3.801737 &&\e^4&&-0.933157 &&\e^5&&+27.3463 &&\e^6, \\
&\vec{J}_{\mathrm{B_6}}(1) &&= && && && && && && && +4.3735 &&\e^4 &&+ 21.281 &&\e^5 &&- 1.94907 &&\e^6, \\
&\vec{J}_{\mathrm{B_7}}(1) &&= +0.25000&&+ 0.640857 &&\e&&- 0.171274&&\e^2&&- 8.80880&&\e^3&&- 76.85069 &&\e^4&&- 609.84213 &&\e^5&&- 3852.6070 &&\e^6, \\
&\vec{J}_{\mathrm{F_1}}(1) &&= -0.319444&&- 1.693693 &&\e&&- 4.661301 &&\e^2&&- 7.989919 &&\e^3&&- 3.492257 &&\e^4&&+ 83.79357&&\e^5&&+804.3009 &&\e^6, \\
&\vec{J}_{\mathrm{G_1}}(1) &&= +0.208333&&+ 1.144388 &&\e&&+ 3.392203 &&\e^2&&+ 8.70991 &&\e^3&&+ 37.4953 &&\e^4&&+ 236.8331 &&\e^5&&+1368.398 &&\e^6, \\
&\vec{J}_{\mathrm{I_2}}(1) &&= +0.333333&&+ 0.183102\, &&\e&&- 7.280536 &&\e^2&&- 20.74864 &&\e^3&&+ 85.02605 &&\e^4&&+ 858.2483 &&\e^5&&+ 3360.3855 &&\e^6.
\end{alignat*}
\end{minipage}
}\\[11pt]
where, for completeness, we added a numerical evaluation also for the non planar family B1, which was already considered in~\cite{Henn:2023vbd}.
The series expansion was truncated at power $n_{\mathrm{max}}=25$ and we report only the digits which are in agreement with the corresponding \texttt{AmFlow} evaluation.
\section{Results}
\label{sec: results}
In this section, we use the canonical master integrals derived in section~\ref{sec: canonical} and solved in terms of iterated integrals in section~\ref{sec: boundaries} to compute the supersymmetric form factor and a part of the $H$+jet amplitude. 
We will demonstrate the beneficial features of the construction of graded transcendental functions in two distinct ways. 
Firstly, it will allow us to express the results efficiently, avoiding cancellations of spurious features. 
Secondly, we will be able to determine all the new functions which appear at this loop order, and hence the analytic structure of the leading-colour $H$+jet amplitudes, completely bypassing an expensive IBP reduction.

We start with the list of canonical combinations for the 9 families derived in this paper, supplemented with the families $A, A_2, A_3, B_1$ considered in~\cite{Henn:2023vbd} and with all planar integrals in the family PL 
as defined in~\cite{Gehrmann:2023zpz}. 
While this set of master integrals covers both quantities we aim to compute, 
not all integrals are independent and a minimal subset could
be identified using shifts in the loop momenta and a further IBP reduction. 
In practice, this step is computationally expensive and unnecessary.
As described above, in our approach we will
detect this minimal set automatically during the construction of the minimal basis of functions.

More explicitly, after applying all 6 kinematic crossings to all master
integrals in the various integral families, we are left with  
a total of 13812 seemingly different canonical combinations. 
The construction of the matrix $S$ defined in eq.~(\ref{eq: Smatrix}) from the Chen iterated solutions reveals that up to $\mathcal{O}(\e^6)$, there are only 1282 unique combinations of functions among all the integrals, a basis of the space $\mathcal{M}_6$.
Next, we grade the minimal basis according to a notion of
complexity inspired by our expectations of the physical properties
of the form factor and the amplitude.
Firstly, we want to discriminate iterated integrals which contain the letters $l_{1-6}$ only, as these are known from lower loop orders and we expect that
the pole structure of any amplitude should be expressible in terms of these functions. 
They will also suffice to express any three-loop quantity that does not require more than the two-loop letters even in its finite remainder, such as the supersymmetric form factors and the maximally transcendental piece of the $Hggg$ amplitude.
It turns out that there are only 93 independent combinations of functions 
involving the letters $l_{7-20}$ across all the 1282 master functions. 
They can be split into three subsets depending on which is the first order where
the new letters appear, either at $\mathcal{O}(\epsilon^4)$, at $\mathcal{O}(\epsilon^5)$ or $\mathcal{O}(\epsilon^6)$.
Based on the observation that the three types of letters identified in table~\ref{tab:alphabet} do not mix up to the $\e$ order relevant for the finite
remainder of the amplitudes, we separate the functions which contain only letters of the parabolic type, those which contain also the hyperbolic type, and finally also the square roots.
In each of the mentioned subsets, we make sure that the elements are linearly independent, and that no combination of elements outside of this subset is equal to them.
In other words, we break $\mathcal{M}_6$ into a direct sum of subspaces, and the functions in each subset are a basis of the subspace they span.

\begin{table}[t]
\setlength{\tabcolsep}{2.5pt}
\centering
\resizebox{\textwidth}{!}{
\begin{tabular}{l|cc|ccc|ccc|ccc}
& $\makecell[c]{\psi_{1}\\-\psi_{3}}$ & $\makecell[c]{\bm{\psi_{4}}\\\bm{-\psi_{9}}}$ & $\makecell[c]{\psi_{10}\\-\psi_{18}}$  & $\makecell[c]{\psi_{19}\\-\psi_{24}}$  & \multicolumn{1}{l|}{$\makecell[c]{\bm{\psi_{25}}\\\bm{-\psi_{36}}}$ } & $\makecell[c]{\psi_{37}\\-\psi_{51}}$ & $\makecell[c]{\psi_{52}\\-\psi_{63}}$ & $\makecell[c]{\bm{\psi_{64}}\\\bm{-\psi_{93}}}$ & $\makecell[c]{\psi_{94}\\-\psi_{111}}$  & $\makecell[c]{\psi_{112}\\-\psi_{1281}}$  & $\bm{{\color{gray}\psi_{1282}}}$\\ \hline
equals FF in sYM & &  & & &  &  & & & & & \cmark \\
violates ~(\ref{eq:old adjacency})&  &  & * & * &  & * & * & & \cmark & &\\
satisfies (\ref{eq:new adjacency1}, \ref{eq:new adjacency2})& \cmark & \cmark & \cmark & \cmark & \cmark & \cmark & \cmark & \cmark & \cmark & \cmark & \cmark\\ \hline
$l_{7-20}$ appears& \cmark&\cmark &\cmark &\cmark &\cmark  &\cmark  &\cmark &\cmark & & &\\
$\bullet$ from $\mathcal{O}(\epsilon^{4})$ & \cmark & \cmark & & &  &  & & & & &\\
$\hookrightarrow$ only parabolic & &\cmark & & &  &  & & & & &\\
$\hookrightarrow$ also roots & \cmark &  & & &  &  & & & & &\\ \hline
$\bullet$ from $\mathcal{O}(\epsilon^{5})$ & &  & \cmark & \cmark & \cmark &  & & & & &\\
$\hookrightarrow$ only parabolic & &  & & &\cmark  &  & & & & &\\
$\hookrightarrow$ also hyperbolic& &  &\cmark &\cmark &  &  & & & & &\\
$\hookrightarrow$ also roots & &  & \cmark & &  &  & & & & &\\ \hline
$\bullet$ from $\mathcal{O}(\epsilon^{6})$& &  & & &  & \cmark & \cmark& \cmark& & &\\
$\hookrightarrow$ only parabolic & & & & &  &  & &\cmark & & &\\
$\hookrightarrow$ also hyperbolic& & & & &  &  \cmark&\cmark & & & &\\
$\hookrightarrow$ also roots & & & & &  &  \cmark & & & & & 
\end{tabular}
}
\caption{An overview of the properties of the trident functions $\psi_{1-1282}$. A check mark indicates that the subset of functions in the column header satisfies the given property and the elements of the subset are linearly independent. Functions marked with an asterisk contain iterated integrals with the letters $l_{7-20}$ and some integrals violating the adjacency condition. Functions which survive in the part of the $H$+jet amplitudes evaluated in this paper are in bold. The form factor is captured by a single trident function (in grey).}
\label{tab:psi overview}
\end{table}

For the well-understood iterated integrals with letters $l_{1-6}$ only, we choose to designate the last function, $\psi_{1282}$, to be equal to the bare form factor equivalent to the bootstrapped result of~\cite{Dixon:2020bbt}.
The remaining functions are again designed such that they cannot replicate the form factor in any linear combination.
This is possible because the form factor must be a combination of our UT master functions with rational numbers as coefficients. 
Such an assignment is convenient for two reasons.
Firstly, if the bootstrapped result is accurate, the reduction of the form factor to our master functions will take a trivial form.
Secondly, we suspect that this function might capture a large portion
of the equivalent $Hggg$ all-plus helicity amplitude. 
The expression for this QCD quantity will therefore take the 
form of a projection onto the $\mathcal{N}=4$ form factor 
and some deviation in the form of the functions $\psi_{1-1281}$. 
Within the remaining set $\psi_{94-1281}$, we isolate the functions which do not obey the conjectured alphabet adjacency conditions on the letters $l_4,l_5,l_6$.
The complete set of functions graded in this way are dubbed \textit{trident functions} $\psi_i$ for their connection to $1\to3$ decays, and their properties are summarised in table~\ref{tab:psi overview}. 
As explained in subsection~\ref{subsec: poles to finite}, we can use this construction to restrict certain iterated integrals from appearing up to the finite part of any quantity expressible in terms of these Feynman integrals, simply based on the fact that spurious terms precede them in the $\e$-expansion of some independent $\psi_i$.
Thousands of complicated functions are eliminated in this way,
\begin{align}
    I_{\mathrm{exc}}=\{
    &I(l_2,l_1,l_4,l_2,l_{3}),
    I(l_1,l_4,l_1,l_5,l_{7}),
    I(l_1,l_2,l_3,l_3,l_{17}), \nonumber
    \\
    &I(l_1,l_4,l_4,l_1,l_{14},l_4),
    I(l_1,l_2,l_4,l_2,l_{10},l_{10}),
    I(l_2,l_2,l_3,l_2,l_{7},l_{8}),\ldots
    \}\,.
\end{align}

Note that the space of 1282 minimal functions and its dimensionality are a feature of the integral topologies at hand.
The subsequent rotations within this set are a matter of choice, informed and justified by expectations based on lower-loop orders and the supersymmetric result.
As more non-planar topologies for this process are computed and new observations or conjectures about the ensuing amplitude are put forward, the grading should be adjusted.
In the supplementary material, we report both the explicit expansions of the trident functions in terms of Chen iterated integrals, as well as a mapping of every canonical basis element to a combination of trident functions with rational numbers as prefactors,
obtained with the help of \texttt{FiniteFieldSolve}~\cite{Mangan:2023eeb}.

\subsection{Direct analytic computation of the $\mathcal{N}=4$ sYM form factor}
\label{sec:eval_phi2}
The $\mathcal{N}=4$ sYM form factor given in eq.~\eqref{eq: form factor definition}  has a perturbative expansion
\begin{align}
    F_{\Tr(\phi^2)} = \sum_L g^{2L} F_{\Tr(\phi^2)}^{(L)} = 
    F^{(0)}_{\Tr(\phi^2)}\sum_L g^{2L} I_{\Tr(\phi^2)}^{(L)}\,, \label{eq:Fexp}
\end{align}
in the coupling
\begin{align}
    g^2 = g^2_{\mathcal{N}=4}\frac{N}{16\pi^2}(4\pi e^{-\gamma_E})^\e\,,
\end{align}
where $g_{\mathcal{N}=4}$ is the bare coupling and $\gamma_E$ is the Euler-Mascheroni constant. 
In eq.~\eqref{eq:Fexp}, $F^{(0)}_{\Tr(\phi^2)}$ represents the tree-level result and
it is the following holomorphic function in spinor variables, proportional to the antisymmetric $\mathrm{SU(N)}$ structure constant,
\begin{align}
    F^{(0)}_{\Tr(\phi^2)} = f^{a_1a_2a_3}\frac{\Braket{12}^2}{\Braket{12}\Braket{23}\Braket{31}}\,.
\end{align}
Having factored out the tree level, the loop corrections are functions of 
 uniform transcendental weight. 

It is well-known that this form factor exponentiates 
according to the Bern-Dixon-Smirnov (BDS) ansatz~\cite{Bern:2005iz,Henn:2011by} and one can
construct a remainder function
\begin{align}
I_{\text{Tr}\left(\phi^{2}\right)} & =I_{\text{Tr}\left(\phi^{2}\right)}^{\text{BDS}}\exp\left(R\right)\,,
\end{align}
where $I_{\text{Tr}\left(\phi^{2}\right)}^{\text{BDS}}$ corresponds to the exponentiated one-loop factor
\begin{align}
   I_{\text{Tr}\left(\phi^{2}\right)}^{\text{BDS}}=\exp \left[\sum_{L=1}^\infty g^{2L}\left(f^{(L)}\left(\epsilon\right)I^{(1)}_{\text{Tr}\left(\phi^{2}\right)}\left(L\epsilon\right)+C^{(L)}\right)\right]\,
\end{align}
with
\begin{align}
    I_{\text{Tr}\left(\phi^{2}\right)}^{\left(1\right)}\left(\e\right)=&-\frac{1}{\e^2}\sum_{i=1}^3\left(\frac{\mu^2}{-s_{i,i+1}}\right)^{\e}-2\left(\text{Li}_2\left(1-x\right)+\text{Li}_2\left(1-y\right)+\text{Li}_2\left(1-z\right)\right)\nonumber\\
    &-\log(x)\log(y)-\log(y)\log(z)-\log(z)\log(x)+\frac{9}{2}\zeta_2\,.
\end{align}
The finite remainder also admits a perturbative expansion in the same coupling constant,
\begin{align}
R & =\sum_{L=2}^{\infty}g^{2L}R^{\left(L\right)}\,.
\end{align}
Explicitly, the two- and three-loop finite remainder functions are
\begin{align}
R^{\left(2\right)} =&+I_{\text{Tr}\left(\phi^{2}\right)}^{\left(2\right)}\left(\e\right)-\frac{1}{2}\left(I_{\text{Tr}\left(\phi^{2}\right)}^{\left(1\right)}\left(\e\right)\right)^{2}-f^{\left(2\right)}\left(\e\right)I_{\text{Tr}\left(\phi^{2}\right)}^{\left(1\right)}\left(2\e\right)-C^{\left(2\right)}+\mathcal{O}\left(\e\right)\,,\\
R^{\left(3\right)}  =&+I_{\text{Tr}\left(\phi^{2}\right)}^{\left(3\right)}\left(\e\right)+\frac{1}{3}\left(I_{\text{Tr}\left(\phi^{2}\right)}^{\left(1\right)}\left(\e\right)\right)^{3}-I_{\text{Tr}\left(\phi^{2}\right)}^{\left(2\right)}\left(\e\right)I_{\text{Tr}\left(\phi^{2}\right)}^{\left(1\right)}\left(\e\right)\nonumber\\
&-f^{\left(3\right)}\left(\e\right) I_{\text{Tr}\left(\phi^{2}\right)}^{\left(1\right)}\left(3\e\right)-C^{\left(3\right)}+\mathcal{O}\left(\e\right)
\label{eq:3l_bds}
\end{align}
with
\begin{align}
f^{\left(1\right)}\left(\e\right) & = 1\,, \nonumber\\
C^{\left(1\right)}& = 0\,,\nonumber\\
f^{\left(2\right)}\left(\e\right) & =-2\zeta_{2}-2\zeta_{3}\e-2\zeta_{4}\e^{2}\,,\nonumber \\
C^{\left(2\right)} & =4\zeta_{4}\,,\nonumber \\
f^{\left(3\right)}\left(\e\right) & =4\left(\frac{11}{2}\zeta_{4}+\left(6\zeta_{5}+\zeta_{2}\zeta_{3}\right)\e+\left(\frac{1909}{48}\zeta_{6}+31\zeta_{3}^{2}\right)\e^{2}\right)\,,\nonumber \\
C^{\left(3\right)} & =16\zeta_{3}^{2}-\frac{181}{3}\zeta_{6}\,.
\end{align}

While $R^{\left(2\right)}$ was calculated analytically leveraging
results for all two-loop Feynman integrals~\cite{Gehrmann:2000zt,Gehrmann:2001ck}
and a unitarity-based construction of the integrand~\cite{Brandhuber:2012vm},
the calculation of $R^{\left(3\right)}$ has only been performed numerically
at a specific phase-space point ($s_{12}=s_{23}=s_{13}=-2$ with
$s_{ij}=2p_{i}\cdot p_{j}$)~\cite{Lin:2021kht,Lin:2021qol}. 
Notably, the bootstrapping approach has extended the calculation of $R$
all the way up to eight loops~\cite{Dixon:2020bbt,Dixon:2022rse}.
These higher-loop calculations exhibit a surprising simplicity in
the functional representation of the remainders, which is expressed exclusively
in terms of two-dimensional  harmonic polylogarithms~\cite{Gehrmann:2001jv}.

The simplicity of the finite remainder for this form
factor becomes fully manifest only when a special normalisation is considered, referred to as the \textit{BDS-like} normalisation.
This choice makes Steinmann relations manifest
and it was crucial for bootstrapping the calculation of the form factor
all the way through eight loops~\cite{Dixon:2020bbt,Dixon:2022rse}.
The form factor in the BDS-normalisation can be expressed as
\begin{align}
I_{\text{Tr}\left(\phi^{2}\right)} & =I_{\text{Tr}\left(\phi^{2}\right)}^{\text{BDS-like}}\times\mathcal{E}\,,
\end{align}
where the relation between $\mathcal{E}$ and $R$ reads
\begin{align}
    \mathcal{E}=\exp\left[\frac{1}{4}\Gamma_\text{cusp}\mathcal{E}^{(1)}+R\right]\,
\end{align}
and the cusp anomalous dimension is
\begin{align}
    \Gamma_{\text{cusp}}=4g^2-8\zeta_2 g^4+88\zeta_4 g^6 -4\left(219 \zeta_6+8\zeta_3^2\right)g^8+\mathcal{O}\left(g^{10}\right)\,.
\end{align}
The finite remainder function $\mathcal{E}$ also admits a perturbative expansion in the coupling constant, 
\begin{align}
    \mathcal{E}=\sum_{L=1}^{\infty} g^{2L}\mathcal{E}^{(L)}
\end{align}
with the one-loop function defined as
\begin{align}
    \mathcal{E}^{(1)}=2\left[\text{Li}_2\left(1-\frac{1}{x}\right)+\text{Li}_2\left(1-\frac{1}{y}\right)+\text{Li}_2\left(1-\frac{1}{z}\right)\right]\,.
\end{align}

For the first time at three loops, the form factor has a leading- and a subleading-colour part,
\begin{align}
I_{\Tr(\phi^{2})}^{(3)}=I_{\Tr(\phi^{2})}^{(3),\mathrm{PL}}+\frac{12}{N^{2}}I_{\Tr(\phi^{2})}^{(3),\mathrm{NPL}}\,.
\end{align}
These contributions are conventionally referred to as \textit{planar}
and \textit{non-planar}, despite both terms receiving
contributions from non-planar diagrams. 
The three-loop integrands
for these quantities were given in terms of Mandelstam invariants
and scalar integrals of rank at most $s=2$ in~\cite{Lin:2021kht,Lin:2021qol}.
In this paper, we focus on the calculation of the planar contribution,
utilising known analytic results for planar integrals along with our
novel calculation of the non-planar integrals $B_1, B_3, B_4, F_1, G_1$,
and $I_2$, as depicted in fig.~\ref{fig: gluonic webs}. 
This leads to the first analytic evaluation of $I_{\Tr(\phi^{2})}^{(3),\mathrm{PL}}$
(and consequently $R^{(3),\mathrm{PL}}$) from a direct computation
of the relevant three-loop Feynman integrals. 

First, we had to extend the one- and two-loop results to transcendental weight 6. 
The integrands were obtained in~\cite{Penante:2014sza,Brandhuber:2012vm} and the master integrals extended to this weight in~\cite{Gehrmann:2023jyv}.
We calculated the one- and two- loop form factors to higher orders in $\e$ and checked against those reported in~\cite{Penante:2014sza} up to the finite part.

For the three-loop calculation, 
we mapped the integrands from~\cite{Lin:2021kht,Lin:2021qol} 
onto a subset of our integral topologies $B_1, B_3, B_4, F_1, G_1, I_2$ and 
the topologies $A, A_2, A_3, E_1, E_2$ defined and evaluated in~\cite{Henn:2023vbd} with the help of \texttt{Reduze}.
Next, we generated IBP identities for all integral families, as well as for 
permutations of external momenta using \texttt{LiteRed}. 
Finally, we solved the large system of linear equations over finite fields using \texttt{FiniteFlow}~\cite{Peraro:2019svx}.
As a result, the form factor can be expressed in terms of a subset of 1399 of the canonical basis elements with rational numbers as prefactors, which is consistent with the fact that amplitudes in $\mathcal{N}=4$ sYM theory are UT.
In terms of trident functions, the reduction of the bare form factor is by definition just
\begin{align}
I^{(3)}_{\Tr(\phi^2)} = \sum_{i=1}^{1281} 0\cdot\psi_{i} + 1\cdot\psi_{1282}\,.
\end{align}
In this form, it is manifest that while several new letters appear in the relevant master integrals (see table~\ref{tab:alphabet}), the resulting form factor is remarkably simple and contains only the six linear letters
\begin{align}
\vec{\alpha}_{\Tr(\phi^{2}),\mathrm{PL}}= & \left\{ x,y,z,1-x,1-y,1-z\right\} \,.
\label{eq:FFalphabet}
\end{align}
This confirms the three-loop result obtained from the bootstrapping
calculation.
The same pattern of cancellation is observed in the explicit analytic calculation of the three-loop BDS-like remainder of the form factor of $\text{Tr}(\phi^3)$~\cite{Henn:2024pki}.

\subsection{Analytic structure of leading-colour $H$+jet amplitudes}
\label{subsec: Hjet results}
We finally move to the QCD amplitudes involving a Higgs boson and three 
strongly interacting partons in the effective theory, see eq.~\eqref{eq:Heft}. 
These are expected to retain some features of the form factor, but are overall 
significantly more complicated due to the mixing of terms of different transcendental 
weight and rational functions.
The partonic channels relevant for $H$+jet production at hadron colliders are the crossings of the two amplitudes $H \rightarrow ggg$ and $H \rightarrow q\bar{q}g$. 
Their loop corrections can be broken down into gauge-invariant terms proportional to different powers of the number of quark colours $N$ and the number of quark flavours propagating in loops $N_f$. 
The terms in the colour expansion are separated by two powers of these parameters. 
Given that they are of similar size (e.g. $N=3$ and $N_f=5$) and $N^2\sim N_f^2$ are of $\mathcal{O}(10)$, we can identify the leading terms in the colour expansion as
\begin{align}
    N^3,\quad N_f N^2,\quad N_f^2N\:\: \mathrm{and} \:\: N_f^3\,.
    \label{eq:colour factors}
\end{align}
In pure QCD, leading-colour terms contain only planar diagrams~\cite{tHooft:1973alw}. 
As was shown in a recent computation, this holds true when considering the production
of an electroweak vector boson and a jet~\cite{Gehrmann:2023zpz}. 

The situation is different in the case of $H$+jet production. In fact, while
the terms proportional to $N_f^2$ and $N_f^3$ still 
contain only very simple planar diagrams (and therefore only letters $l_{1-6}$), the first two colour factors in~\eqref{eq:colour factors} 
correspond to planar three-loop $ggg$
and $q\bar q g$ webs with a Higgs boson connecting to any gluon (internal or
external) 
in the web. Consequently, they include 
planar and non-planar diagrams, and the 
class of transcendental functions which describes them is yet unknown.

The independent helicity amplitudes of the gluonic and quark-antiquark channels can be expressed in terms of the scalar coefficients $\alpha, \beta$ and $\gamma$, which admit the following  perturbative expansion in the strong coupling $\alpha_{s}$,
\begin{alignat}{3}
A_{ggg}^{+++}&=\mathcal{C}\, \frac{m^4  f^{a_1 a_2 a_3}}{\langle12\rangle\langle23\rangle\langle31\rangle} &&\left[-1+\left(\frac{\alpha_s}{2\pi}\right)\alpha^{(1)}+\left(\frac{\alpha_s}{2\pi}\right)^2\alpha^{(2)}+\left(\frac{\alpha_s}{2\pi}\right)^3\alpha^{(3)}+\mathcal{O}(\alpha_s^4)\right]\,,\nonumber\\    
    A_{ggg}^{++-}&=\mathcal{C}\, \frac{[12]^3 f^{a_1 a_2 a_3}}{[23][13]}&&\left[-1+\left(\frac{\alpha_s}{2\pi}\right)\beta^{(1)}+\left(\frac{\alpha_s}{2\pi}\right)^2\beta^{(2)}+\left(\frac{\alpha_s}{2\pi}\right)^3\beta^{(3)}+\mathcal{O}(\alpha_s^4)\right]\,,\nonumber\\
A_{q\bar{q}g}^{LR+}&=\mathcal{C}\, \frac{[23]^3\, T^{a_3}_{i,j}}{[12]}&&\left[+1+\left(\frac{\alpha_s}{2\pi}\right)\gamma^{(1)}+\left(\frac{\alpha_s}{2\pi}\right)^2\gamma^{(2)}+\left(\frac{\alpha_s}{2\pi}\right)^3\gamma^{(3)}+\mathcal{O}(\alpha_s^4)\right]\,.
\end{alignat}
with $\mathcal{C} = \frac{\lambda \sqrt{4\pi\alpha_s}}{\sqrt{2}}$.
A detailed derivation of the independent helicity structures and the terms of the expansions up to 2 loops were given in~\cite{Gehrmann:2023jyv}. 
In this work, we showcase how the grading of transcendental functions allows us to evaluate parts of the finite remainders of the three-loop scalar coefficients $\alpha^{(3)}$, $\beta^{(3)}$, $\gamma^{(3)}$, in particular those which carry a dependence on the new three-loop letters.

To achieve this, we take advantage of the fact that these new letters appear only in diagrams with 8 or more internal propagators, pictured in fig.~\ref{fig: sources}. 
In practice, we compute the contribution of all Feynman diagrams to the three-loop $H$+jet integrand in \texttt{FORM}~\cite{Vermaseren:2000nd} using standard multiloop methods
and set to zero all subtopologies with 7 or fewer lines. 
This drastically reduces the complexity of the IBP reduction, but it allows us to retain full dependence on the functions with new kernels.
As our goal is to determine the space of surviving functions, 
not evaluate or approximate the amplitude, we opted to perform the reduction for several numerical values of $x,y$, keeping $\e$ symbolic.
For reference, in the supplementary material we provide the coefficients of all 
functions with new alphabet letters appearing in the colour factors listed in eq.~\eqref{eq:colour factors} of the helicity amplitudes $\alpha^{(3)}$, $\beta^{(3)}$, $\gamma^{(3)}$ at the arbitrary point $x=3/13, y=11/17$. 

To illustrate the significance of our results, we will first discuss in detail the coefficient $\alpha^{(3)}$.
The finite remainder of the all-plus helicity amplitude of $Hggg$, written in terms of the trident functions $\psi_{1-93}$, reads 
\begin{align}
    \alpha^{(3)}|_{N^3} = \e^{-6}\,\Bigl[&
    + \frac{1746}{48841} \e^2\, \psi_{4} 
    + \frac{22}{289} \e^2\, \psi_{6}
    + \frac{10}{169} \e^2\, \psi_{7} \nonumber\\
    &+\frac{11}{18}\Bigl( \frac{37}{3} \e\: \psi_{25}
    + 5 \e\: \psi_{26}
    + 6 \e\: \psi_{27}
    + \frac{11}{2} \e\: \psi_{28}
    + 8 \e\: \psi_{29} \nonumber\\
    &- 7 \e\: \psi_{30} 
    - 5 \e\: \psi_{31}
    - \frac{5}{2} \e\: \psi_{32}
    + 2 \e\: \psi_{33}
    - \frac{22}{3} \e\: \psi_{34}
    + 2 \e\: \psi_{35}
    \Bigl)\Bigl]
    \nonumber\\
    + &\mathrm{\:terms\:with\:letters\:} l_{1-6} \mathrm{\:only}\nonumber\\
    + &\mathcal{O}(\e)\,.
\end{align}
A number of properties can be immediately read off from the expression.
First of all, none of the trident functions with new letters starting at $\mathcal{O}(\e^6)$, that is $\psi_{37-93}$, appear in the finite part.
Similarly, the trident functions $\psi_{1-3}$ and $\psi_{10-24}$ with hyperbolic or square root letters are absent.
Further, the coefficients of trident functions containing parabolic letters at $\mathcal{O}(\epsilon^{4})$, i.e. $\psi_{4-9}$, are multiplied by coefficients of at least $\mathcal{O}(\epsilon^2)$, excluding them from the poles of the amplitude.
Equivalently, trident functions with parabolic letters at $\mathcal{O}(\epsilon^{5})$ are multiplied by coefficients of at least $\mathcal{O}(\epsilon)$.
These findings confirm the logic presented in subsection~\ref{subsec: poles to finite}: 
the knowledge of the IR and UV poles (in this case simply the absence of $l_{7-20}$) can restrict the $\epsilon$ expansions of the coefficients of the trident functions.
Referring to the properties listed in table~\ref{tab:psi overview}, the adjacency conditions of eqs.~\eqref{eq:old adjacency} and~\eqref{eq:new adjacency1}, \eqref{eq:new adjacency2} are also manifestly satisfied.
Considering the small number of $\psi$ functions involved in this expansion and the simplicity of their coefficients compared to the equivalent in traditional MIs, it is clear that our basis of transcendental functions is closely aligned with the subspace in which the amplitude resides.
For this reason, we expect the future IBP reduction of the full amplitude 
to this basis to be significantly easier, especially if employing finite-fields based
techniques which aim at reconstructing only the rational functions appearing in the final
result.

The simplicity of the rational numbers in the coefficients
is evocative of the underlying rational functions.
In fact, starting from a handful of numerical evaluations, we were able to reconstruct the exact analytic form of the corresponding rational functions and
insert the explicit expressions for the trident functions $\psi_{1-93}$ with new letters
in terms of Chen iterated integrals.
The result can be written in an extremely compact form as follows
{
\setlength{\abovedisplayskip}{5pt}
\setlength{\belowdisplayskip}{5pt}
\begin{align}
    \alpha^{(3)} = 2N^2\beta_0\Bigl\{&\left[f_5(x,y)-f_1(x,y)f_4(x,y)\right]\times S_3\nonumber\\
    +\zeta_2&\left[f_{3a}(x,y)+f_{3b}(z,y)-f_1(x,y)f_{2a}(x,y)-f_1(z,y)f_{2b}(z,y)\right]\times P_{x\leftrightarrow y}\Bigl\}\nonumber\\
    -\frac{1}{3}N^2(N-N_f)\Bigl\{&\left[(1-y)y f_4(x,y)\right]\times S_3\nonumber\\
    +\zeta_2&\left[(1-y)y (f_{2a}(x,y)+f_{2b}(z,y))\right]\times P_{x\leftrightarrow y}\Bigl\}\nonumber\\
    + &\mathrm{\:terms\:with\:letters\:} l_{1-6} \mathrm{\:only.}
    \label{eq: alpha3 new functions}
\end{align}
}
In eq.~\eqref{eq: alpha3 new functions}, $\beta_0$ is the first coefficient
of the QCD beta function, the symbol $\times S_3$ indicates all 6 permutations
of the three external legs, while $P_{x\leftrightarrow y}$ is only the swap $x\leftrightarrow y$, i.e. $p_2\leftrightarrow p_3$. The functions $f_i(x,y)$ are simple combinations of Chen iterated integrals:
\begin{align}
    f_1(x,y) = &+ I(\lambda)\\
    f_{2b}(x,y) =&+2I(\lambda,\bm{l_{11}})\\
    f_{3b}(x,y) =&+4I(\lambda,\lambda,\bm{l_{11}})\\
    f_{2a}(x,y) =&+I(\kappa,\bm{l_{11}})\\
    f_{3a}(x,y) = &+I(\lambda,\kappa,\bm{l_{11}})+I(\kappa,\lambda,\bm{l_{11}})\\
    f_4(x,y) =  +2[&+I(l_2,\mu,\lambda,\bm{l_{11}})-I(l_3,\nu,\lambda,\bm{l_{11}})]\nonumber\\
    &
    -2I(l_2,l_2,\kappa,\bm{l_{11}})+I(l_3,l_6,\kappa,\bm{l_{11}}) \\
    f_5(x,y) = +4[&+I(l_2,\mu,\lambda,\lambda,\bm{l_{11}})
    -I(l_3,\nu,\lambda,\lambda,\bm{l_{11}})]\nonumber\\
    +2[&+I(l_2,\lambda,\mu,\lambda,\bm{l_{11}})
    -I(l_3,\lambda,\nu,\lambda,\bm{l_{11}})]\nonumber\\
    +2[&+I(\lambda,l_2,\mu,\lambda,\bm{l_{11}})
    -I(\lambda,l_3,\nu,\lambda,\bm{l_{11}})]\nonumber\\
    &  
    -2I(l_2,l_2,\kappa,\lambda,\bm{l_{11}})
    +I(l_3,l_6,\kappa,\lambda,\bm{l_{11}})\nonumber\\
    &-2I(l_2,l_2,\lambda,\kappa,\bm{l_{11}})
    +I(l_3,l_6,\lambda,\kappa,\bm{l_{11}})\nonumber\\
    &-2I(l_2,\lambda,l_2,\kappa,\bm{l_{11}})
    +I(l_3,\lambda,l_6,\kappa,\bm{l_{11}})\nonumber\\
    &-2I(\lambda,l_2,l_2,\kappa,\bm{l_{11}})
    +I(\lambda,l_3,l_6,\kappa,\bm{l_{11}})
\end{align}
written in terms of the conveniently combined letters 
\begin{alignat}{4}
    \lambda&\equiv\lambda (x,y) &&= l_1 \cdot l_2^{-1} \cdot l_5^{-1} &&= \frac{x}{y(1-y)}\,,\nonumber\\
    \kappa&\equiv\kappa (x,y) &&= l_3 \cdot l_5^{-2} &&= \frac{1-x-y}{(1-y)^2}\,, \nonumber\\
    \mu&\equiv\mu(x,y) &&= l_3 \cdot l_5^{-1} &&= \frac{1-x-y}{1-y}\,, \nonumber\\
    \nu&\equiv\nu(x,y) &&= l_6 \cdot l_2^{-1} &&= \frac{x+y}{y}\,, 
    \label{eq:lambda def}
\end{alignat}
which have the following property
\begin{align}
    \lambda(x,y)=1\quad&\Longleftrightarrow \quad\bm{l_{11}}=0\,,\nonumber\\
    \kappa(x,y)=1\quad&\Longleftrightarrow \quad\bm{l_{11}}=0\,,\nonumber\\
    \mu(x,y)=1\quad&\Longleftrightarrow \quad l_{1}=0\,,\nonumber\\
    \nu(x,y)=1\quad&\Longleftrightarrow \quad l_{1}=0\,.
\end{align}

After applying the permutations $S_3$ and $P_{x\leftrightarrow y}$ in eq.~\eqref{eq: alpha3 new functions}, the five remaining kinematic crossings of eq.~\eqref{eq:lambda def} will also appear, representing other combinations of the letters $l_{1-6}$ and matching the singularity of another crossing of $\bm{l_{11}}$ or $l_1$.
For later, we also define the combination
\begin{align}
    \sigma(x,y) = l_1\cdot l_5^{-1}= \frac{x}{1-y}\,,
    \label{eq:nu def}
\end{align}
for which $\sigma(x,y)=1$ when $l_{3}=0$.

The all-plus helicity amplitude for the decay to three gluons should be symmetric in the gluon momenta $p_1,p_2,p_3$.
This property is manifestly satisfied at symbol level and in all terms which do not involve transcendental constants (see the first and third lines of eq.~\eqref{eq: alpha3 new functions}).
However, the choice of the base point for the iterated integrals $(x=0,y=0,z=1)$ introduces an asymmetry in the coefficients of the higher-weight transcendental constants in the regularised boundary values at this point.
Consequently, in terms involving $\zeta_2$, the full exchange symmetry appears only as invariance under $x\leftrightarrow y$.
Note that eq.~\eqref{eq: alpha3 new functions} represents both the bare amplitude as well as the IR- and UV-subtracted finite remainder (the difference  consists of iterated integrals containing only the letters $l_{1-6}$ and these integrals were discarded in our analysis).
It is remarkable that the highest-weight part of the expression containing the new three-loop letters, which is of weight five, is proportional to the QCD $\beta_0$. 
These terms are absent in the corresponding supersymmetric quantity 
where the $\beta$ function is zero.
The remaining contribution, with functions of weight 4 and lower, vanishes for $N_f = N$.

Besides the specific properties of the all-plus configuration discussed above, 
our results for the helicity coefficients have the following properties which can be inferred by
direct inspection of the analytic expressions:
\begin{itemize}
    \item The square-root letters $l_{7,8}$ and hyperbolic letters $l_{9-10,17-20}$ drop out of the finite remainders, despite the fact that master integrals which contain them are present.
    \item For all contributions to the $Hggg$ amplitudes, no new letters appear in weight-6 functions, in accordance with the maximum weight conjecture. 
    It is striking that functions of weight 4 and 5 with all six parabolic letters $l_{11-16}$ are still present.
    Whether the weight-6 functions with only two-loop letters also match those of the $\mathcal{N}=4$ form factor
    will be established in a future direct calculation, but if this were the case, the reduction of the amplitude in
    terms of the trident functions would read
    \begin{align}
    \alpha^{(3)}|_{N^3} = 1\cdot\psi_{1282} +  \sum_{i=1}^{1281}c_i\psi_{i}\,,\qquad c_i=\mathcal{O}(\epsilon)\,.
    \end{align}
    \item The $Hgq\bar{q}$ amplitude has no supersymmetric counterpart and already at lower loops exhibits fewer symmetries. Indeed, we observe the appearance of weight-6 functions with the new parabolic letters. More precisely, we find that only the two letters $l_{12}$ and $l_{13}$ contribute.
    \item The new adjacency conditions from the G2 cluster algebra were already satisfied on the level of master integrals. It turns out that also the adjacency conditions of eq.~(\ref{eq:old adjacency}) are satisfied for all the surviving functions with new integration kernels. 
    We expect that these observations will have significant implications for bootstrapping approaches in the future.
    \item Only very few unique rational coefficients appear in the amplitudes, and they are very simple, 
    which is a vindication of the basis of trident functions. 
    Furthermore, we notice that the new quadratic letter only ever appears as the fourth integration kernel. 
    In other words, functions of weight 5 or 6 with the parabolic letter as the fifth or 
    sixth integration kernel drop out.
    We can extend this fourth-entry observation also to terms involving transcendental constants by taking into account the weight of the constant multiplying the iterated integral.
    In other words, terms like $\pi^2I(l_i,l_{11}, l_j, l_k)$ or $\zeta_3 I(l_{11}, l_j, l_k)$ with $i,j,k\in\{1,\ldots,6\}$ also satisfy the condition.
    The iterated integrals multiplied by $\zeta_3$ have a discontinuity corresponding to the parabolic letter and we do not observe any such terms.
    Terms with $\pi^4$ or $\zeta_5$ and a new integration kernel would be completely excluded by the fourth-entry condition and indeed, they do not appear.
\end{itemize}
In conclusion, this analysis allows us to determine the full alphabet for leading-colour $H$+jet amplitudes 
in the decay region as
\begin{align}
\vec{\alpha}_{H\to3\mathrm{partons},\mathrm{L.C.}}=  \Bigl\{&x,y,z,1-x,1-y,1-z,x^2-x+y,
y^2-y+x,\nonumber\\
&\left(x+y\right)^2-x,
-y+\left(1-x\right)^2,
\left(x+y\right)^2-y,
\left(y-1\right)^2-x\Bigl\} \,.
\label{eqs: alphabet_HJCL}
\end{align}
However, after analytic continuation to scattering kinematics, some of the hyperbolic 
letters will also appear, as discussed in section~\ref{subsec: alphabet}.

Now that we have established the relevant function space, we can look for an explicit functional representation,
which will be relevant especially in view of numerical evaluation. 
To this end, it would be convenient if the 
quadratic letter appeared only as the outermost integration kernel of each iterated integral. 
Since the quadratic letters already appear only as the fourth integration kernel, this
form can easily be achieved by exploiting shuffle algebra.
We encounter the following types of integrals:\footnote{Recall our convention for iterated integrals, as defined in eq.~\eqref{eq:Chen_int_f}: the right-most index of the iterated integral corresponds to the outermost integration.}
\begin{align}
\lbrace I(\vec{\omega}_{i_1\ldots i_n}, \omega_q; \vec{x}),I(\vec{\omega}_{i_1\ldots i_n}, \omega_q,\omega_j; \vec{x}),I(\vec{\omega}_{i_1\ldots i_n}, \omega_q,\vec{\omega}_{jk}; \vec{x})\rbrace\,,
\label{eqs: iterated to unshuffle}
\end{align}
where we indicate with $\omega_q$ the quadratic integration kernel, with $\vec{\omega}_{i_1\ldots i_n}$ a sequence of $n$ differential forms $\omega_{i_1}\ldots \omega_{i_n}$ and with $\vec{\omega}_{jk}$ some two linear kernels $\omega_j, \omega_k\in\{l_1,\ldots,l_6\}$. 
By applying the shuffle product~\eqref{eqs: shuffle product}, we can always move $\omega_q$ to last position at the price of introducing products 
of lower-weight iterated integrals in the final expression.
For integrals with one integration kernel after the quadratic one,
\begin{align}
 I(\vec{\omega}_{i_1\ldots i_n}, \omega_q,\omega_j; \vec{x}) =&
 I(\vec{\omega}_{i_1\ldots i_n}, \omega_q;\vec{x}) I(\omega_j;\vec{x}) -\sum_{\vec{\omega}_s=\vec{\omega}_{i_1\ldots i_n} \shuffle\,\omega_j}I(\vec{\omega}_s,\omega_q;\vec{x})\,,
 \label{eqs: shuffle5}
\end{align}
while in the case of two integration kernels after the quadratic one,
\begin{align}
 I(\vec{\omega}_{i_1\ldots i_n}, \omega_q,\vec{\omega}_{jk}; \vec{x}) =& \,
 I(\vec{\omega}_{i_1\ldots i_n}, \omega_q;\vec{x}) I(\omega_{jk};\vec{x}) \nonumber\\
 &-\sum_{\vec{\omega}_s=\vec{\omega}_{i_1\ldots i_n} \shuffle\,\vec{\omega}_{jk}}I(\vec{\omega}_s,\omega_q;\vec{x})\nonumber\\ &-\sum_{\vec{\omega}_s=\vec{\omega}_{i_1\ldots i_n} \shuffle\,\omega_{j}}I(\vec{\omega}_s,\omega_q,\omega_k;\vec{x})\,.
 \label{eqs: shuffle6}
\end{align}
The third line in eq.~\eqref{eqs: shuffle6} still contains unwanted terms, which can again be removed with eq.~\eqref{eqs: shuffle5}.

Starting from this representation, one can specify the path $\gamma$ in eq.~\eqref{eqs:Chen_int} to obtain 
an explicit representation of the integrals. For example, to reach the point $\vec{x}=(x,y)$, one can combine 
two segments, $\gamma_1 \cup \gamma_2$, with
\begin{align}
\gamma_1(t) &= (0,t) \quad \quad t \in [0,y]\,, \nonumber \\
\gamma_2(t) &= (t,y) \quad \quad t \in [0,x]\,.
\label{eqs: fibration}
\end{align}
This choice corresponds to a representation in terms of multiple polylogarithms
with a particular fibration basis.
If we limit ourselves to the alphabet letters linear in the variable~$x$ 
($l_{1-6}$ and 2 out of 6 parabolic letters), this fibration does not introduce square roots 
in the index of the MPLs.
Nonetheless, we have seen that the alphabet for the leading-colour $H$+jet amplitudes involves all the quadratic parabolic letters at once, and a single choice of fibration would introduce indices with square roots. 
We can avoid this at the price of writing the expression in terms of functions which are 
not linearly independent. In fact, parabolic letters can be always written in a form linear in one variable,
\begin{align}
z_1-z_2(1-z_2)\,,
\label{eqs: linearised parab}
\end{align}
where $z_1,z_2$ are two of the three kinematic variables $\{x,y,z\}$. 
Starting from the Chen iterated integral representation obtained by the application of~\eqref{eqs: shuffle5} and~\eqref{eqs: shuffle6}, one can express each individual iterated integral in terms of those two kinematic variables which make the parabolic letter explicitly linear in one of the variables.

Since we only reconstructed the part of the amplitudes containing the quadratic letters, neglecting contributions with linear kernels only, our result is not an integrable combination of iterated integrals.
Nevertheless, it is possible to consider all the possible MPLs expressions with indices $\{0,1\}$ 
if the argument is $z_2$ and with indices $\{0,1,1-z_2,-z_2,z_2(1-z_2)\}$ if the argument is $z_1$,
which could 
correspond to the given iterated integral.  
Next, we can solve for the combination whose symbol correctly reproduces the iterated integral representation with quadratic letters, modulo terms with a purely linear alphabet.
In practice, we individually consider terms proportional to different transcendental constants and, following the ansatz approach, we are able to find an equivalent MPL expression for each of them. 

This procedure allowed us to further investigate the analytic structure of each transcendental function in view of phenomenological applications. 
In the Euclidean region, all master integrals for this process must be real and single-valued. 
However, this does not imply that each of transcendental function is independently real.
Any MPL 
characterised by the linear alphabet $\lbrace l_1, \ldots, l_6 \rbrace$  of the two-loop and the three-loop 
planar case is explicitly real and single valued in the decay region. 
This is not the case when alphabet 
letters $\lbrace l_7, \ldots, l_{20} \rbrace$ are present. 

Interestingly, working at the level of the amplitudes, we found that a rational change of variables exists that makes each individual MPL real and single valued in the decay region. 
In fact, of the set of arguments used to generate an ansatz, only the subset 
$\{0, 1 - z_2, -z_2,z_2(1-z_2)\}$ appears in the result, that is, the letter $1-z_2$ is not present in the surviving MPLs which also have the quadratic index. 
Then, the rational change of variables corresponding to the combined symbol letters defined in eqs.~\eqref{eq:lambda def} and~\eqref{eq:nu def},
\begin{align}
\left\{ \tilde{z}_1 = \sigma(z_1,z_2),   \:\tilde{z}_2 = \lambda(z_1,z_2)\right\},
\label{eqs: change of variables}
\end{align}
maps this restricted set of indices into a linear one. 
Choosing a fibration basis in which $\tilde{z}_1$ is the argument, the new indices of the MPLs in the amplitude are 
\begin{align}
\{0,1,\tilde{z}_2,1+\tilde{z}_2\}\,,
\end{align}
where $0 < \tilde{z}_2 < \infty$ and the inequality $ \tilde{z}_1 < \min\{1, \tilde{z}_1 \}$ holds. The letter $1-z_2$, which decouples from quadratic MPLs, is the only one that would not transform linearly under the change of variables in eq.~\eqref{eqs: change of variables}. 

In conclusion, the targeted numerical IBP reductions of the leading-colour amplitudes for $H\rightarrow ggg$ 
and $H\rightarrow q\bar{q}g$ allowed us to establish the analytic complexity of the final result 
beyond the already known linear alphabet. We identified the most complicated iterated integrals in the amplitudes and we showed how to convert them into MPLs with linear or at most quadratic indices. Finally, we presented a change of variables that makes all appearing MPLs manifestly real in the decay region. 

\section{Conclusion}
\label{sec: conclusion}
In this work, we presented the full set of three-loop master integrals relevant to the evaluation of the form factor of the operator $\Tr(\phi^2)$ in planar $\mathcal{N}=4$ sYM theory and to QCD amplitudes for $H$+jet production in the leading-colour approximation and with $m_t\to\infty$.
We derived a set of canonical differential equations for each contributing integral family, determining the analytic structure of the resulting integrals.
Three new types of alphabet letters appear from $\mathcal{O}(\e^{4})$ in functions of weight 2 and higher.
One type contains a square root which cannot be rationalized without introducing higher powers of invariants in other letters.

The process of deriving canonical bases and fixing boundary constants was described in detail. 
For the latter, we parametrized the differential equations on carefully chosen one-dimensional slices,
on which the alphabet simplifies. Separately on each segment, we solved the simplified DEs 
and enforced the relevant regular or singular boundary behaviour. 
Through a matching of the solutions at the intersections of the segments, we were then 
able to transport all the boundary information to a single kinematic point, 
where the value of all integrals is completely fixed.
All master integrals were solved up to the last $\e$ order contributing 
to a three-loop amplitude in terms of Chen iterated integrals.

Building upon the concept of a minimal graded basis of transcendental functions~\cite{Chicherin:2020oor,Chicherin:2021dyp}, 
we formulated and implemented a method for constructing such a basis tailored to a scattering amplitude. 
Instead of conventional master integrals, we expressed the $\mathcal{N}=4$ sYM form factor in a basis of functions of uniform weight where its remarkable properties are manifestly satisfied. 
In particular, we determined for the first time with a direct calculation its alphabet and 
confirmed several conjectured adjacency conditions. 
We expressed the finite remainder in terms of  Chen iterated integrals as well as MPLs, 
including terms beyond the symbol, and checked agreement with the bootstrapped expression as 
well as the numerical evaluation at a specific kinematic point available in the literature. 

Based on the knowledge of the simplifications that occur in the
$\mathcal{N}=4$ sYM form factor, we devised a function basis for the QCD amplitude.
This graded basis arranges the functions into subsets based on the occurrence of 
various classes of letters at different weights. 
In this basis, it became obvious that any new letter would only appear once
in any iterated integral and several functions would have to be multiplied by
powers of $\e$ to satisfy pole cancellation. Since the new letters only appear 
in integrals with 8 or more lines, we ran multiple numerical IBP reductions with exact dependence on 
$\e$ but
ignoring all integrals with fewer propagators. 
This economical approach allowed us to completely determine the 
part of the QCD amplitude depending on the new set of letters.
Remarkably, we were able to write the amplitude in a form where manifestly all 6 hyperbolic letters 
and all square roots drop out.
The remaining quadratic letters appear only in the $Hgq\bar{q}$ channel, and in lower-than-maximal weight functions in the finite part of the $Hggg$ channel.
As~conjectured, at the symbol level, the alphabet of the QCD amplitude and the related 
$\mathcal{N}=4$ sYM 
form factor coincide.
Finally, we also demonstrated how to express all iterated integrals depending on the new letters
in terms of \emph{explicitly real} MPLs with quadratic letters, paving the way for their phenomenological implementation.

\section*{Acknowledgments}
We would like to thank Antonela Matija\v{s}i\'{c}, Christoph Nega, Kay Sch\"{o}nwald, Vasily Sotnikov and Chen-yu Wang for stimulating discussions. 
We are grateful to the authors of~\cite{Dixon:2020bbt} for providing polylogarithmic expressions for checks and to the authors of~\cite{Aliaj:2024zgp} for useful clarifications. 
This work was supported in part by the Munich Institute for Astro-, Particle and BioPhysics (MIAPbP) and
the Excellence Cluster ORIGINS funded by the Deutsche Forschungsgemeinschaft (DFG, German Research Foundation) under Germany’s Excellence Strategy – EXC-2094-390783311,
by the European Research Council (ERC) under the European Union’s research and innovation programme grant agreements 949279 (ERC Starting Grant HighPHun) and 101019620 (ERC Advanced Grant TOPUP),
and by the Leverhulme Trust, LIP-2021-01.

\appendix
\section{Integral families}\label{app: A}
{
\setlength{\abovedisplayskip}{3pt}
\setlength{\belowdisplayskip}{3pt}
In this appendix, we give the list of denominators and numerators appearing in each integral family, as defined in eq.~\eqref{eq:feynman integral definition}.
We use the shorthand $p_{ij} = p_{i}+p_{j}$ and $p_{ijk} = p_i+p_j+p_k$.
For family $B_1$, 
\begin{alignat*}{10}
P_1&=k_{1}\quad&&\:P_4&&=k_{2}+p_{12}\quad&&\:P_7&&=k_{1}+p_{1}\quad&&\:P_{10}&&=k_{3}-p_{3}\quad&&\:N_3&&=k_{2}-p_{3}\\[-1mm]
P_2&=k_{1}+p_{12}\quad&&\:P_5&&=k_{3}\quad&&\:P_8&&=k_{1}-k_{2}\quad&&\:N_1&&=k_{1}-p_{3}\quad&&\:N_4&&=k_{3}+p_{1}\\[-1mm]
P_3&=k_{2}\quad&&\:P_6&&=k_{2}-k_{3}+p_{123}\quad&&\:P_9&&=k_{2}-k_{3}\quad&&\:N_2&&=k_{2}+p_{1}\quad&&\:N_5&&=k_{1}-k_{3}\,.
\end{alignat*}
The definition of family $B_3$ is
\begin{alignat*}{10}
P_1&=k_{1}\quad&&\:P_4&&=k_{1}-k_{2}+k_{3}-p_{123}\quad&&\:P_7&&=k_{1}-k_{2}\quad&&\:P_{10}&&=k_{3}-p_{12}\quad&&\:N_3&&=k_{2}-p_{1}\\[-1mm]
P_2&=k_{1}-p_{123}\quad&&\:P_5&&=k_{3}\quad&&\:P_8&&=k_{2}-k_{3}\quad&&\:N_1&&=k_{1}-k_{3}\quad&&\:N_4&&=k_{1}-p_{2}\\[-1mm]
P_3&=k_{2}\quad&&\:P_6&&=k_{3}-p_{123}\quad&&\:P_9&&=k_{3}-p_{1}\quad&&\:N_2&&=k_{1}-p_{1}\quad&&\:N_5&&=k_{2}-p_{2}\,.
\end{alignat*}
The definition of family $B_4$ is
\begin{alignat*}{10}
P_1&=k_{1}\quad&&\:P_4&&=k_{1}-k_{2}\quad&&\:P_7&&=k_{2}-k_{3}\quad&&\:P_{10}&&=k_{3}-p_{3}\quad&&\:N_3&&=k_{1}-k_{3}\\[-1mm]
P_2&=k_{1}+p_{1}\quad&&\:P_5&&=k_{1}-k_{2}+p_{3}\quad&&\:P_8&&=k_{2}-p_{3}\quad&&\:N_1&&=k_{2}\quad&&\:N_4&&=k_{2}+p_{1}\\[-1mm]
P_3&=k_{1}+p_{12}\quad&&\:P_6&&=k_{2}+p_{12}\quad&&\:P_9&&=k_{3}+p_{12}\quad&&\:N_2&&=k_{3}\quad&&\:N_5&&=k_{3}+p_{1}\,.
\end{alignat*}
The definition of family $B_5$ is
\begin{alignat*}{10}
P_1&=k_{1}\quad&&\:P_4&&=k_{3}-p_{1}\quad&&\:P_7&&=k_{2}-k_{3}\quad&&\:P_{10}&&=k_{2}+p_{3}\quad&&\:N_3&&=k_{1}-p_{12}\\[-1mm]
P_2&=k_{2}\quad&&\:P_5&&=k_{3}-p_{12}\quad&&\:P_8&&=k_{1}-k_{2}-p_{3}\quad&&\:N_1&&=k_{1}-p_{1}\quad&&\:N_4&&=k_{1}-k_{3}\\[-1mm]
P_3&=k_{3}\quad&&\:P_6&&=k_{1}-k_{2}\quad&&\:P_9&&=k_{2}-k_{3}+p_{123}\quad&&\:N_2&&=k_{2}-p_{1}\quad&&\:N_5&&=k_{2}-p_{12}\,.
\end{alignat*}
The definition of family $B_6$ is
\begin{alignat*}{10}
P_1&=k_{1}\quad&&\:P_4&&=k_{3}-p_{1}\quad&&\:P_7&&=k_{2}-k_{3}\quad&&\:P_{10}&&=k_{1}-k_{3}\quad&&\:N_3&&=k_{1}-p_{12}\\[-1mm]
P_2&=k_{2}\quad&&\:P_5&&=k_{3}-p_{12}\quad&&\:P_8&&=k_{2}-k_{3}+p_{123}\quad&&\:N_1&&=k_{1}-p_{1}\quad&&\:N_4&&=k_{1}-k_{2}-p_{3}\\[-1mm]
P_3&=k_{3}\quad&&\:P_6&&=k_{1}-k_{2}\quad&&\:P_9&&=k_{2}+p_{3}\quad&&\:N_2&&=k_{2}-p_{1}\quad&&\:N_5&&=k_{2}-p_{12}\,.
\end{alignat*}
The definition of family $B_7$ is
\begin{alignat*}{10}
P_1&=k_{1}\quad&&\:P_4&&=k_{1}-p_{1}\quad&&\:P_7&&=k_{1}-p_{12}\quad&&\:P_{10}&&=k_{2}-k_{3}\quad&&\:N_3&&=k_{2}+p_{3}\\[-1mm]
P_2&=k_{2}\quad&&\:P_5&&=k_{2}-p_{1}\quad&&\:P_8&&=k_{3}-p_{12}\quad&&\:N_1&&=k_{1}-k_{2}-p_{3}\quad&&\:N_4&&=k_{1}-k_{3}\\[-1mm]
P_3&=k_{3}\quad&&\:P_6&&=k_{3}-p_{1}\quad&&\:P_9&&=k_{1}-k_{2}\quad&&\:N_2&&=k_{2}-k_{3}+p_{123}\quad&&\:N_5&&=k_{2}-p_{12}\,.
\end{alignat*}
The definition of family $F_1$ is
\begin{alignat*}{10}
P_1&=k_{1}-k_{3}\quad&&\:P_4&&=k_{2}+p_{12}\quad&&\:P_7&&=k_{1}-k_{2}\quad&&\:P_{10}&&=k_{3}-p_{3}\quad&&\:N_3&&=k_{1}-p_{3}\\[-1mm]
P_2&=k_{1}+p_{1}\quad&&\:P_5&&=k_{2}-k_{3}+p_{123}\quad&&\:P_8&&=k_{3}\quad&&\:N_1&&=k_{3}+p_{12}\quad&&\:N_4&&=k_{1}\\[-1mm]
P_3&=k_{1}+p_{12}\quad&&\:P_6&&=k_{2}-k_{3}\quad&&\:P_9&&=k_{3}+p_{1}\quad&&\:N_2&&=k_{2}+p_{1}\quad&&\:N_5&&=k_{2}\,.
\end{alignat*}
The definition of family $G_1$ is
\begin{alignat*}{10}
P_1&=k_{1}-k_{3}\quad&&\:P_4&&=k_{1}-k_{2}+p_{123}\quad&&\:P_7&&=k_{1}-k_{2}\quad&&\:P_{10}&&=k_{3}-p_{3}\quad&&\:N_3&&=k_{1}-p_{3}\\[-1mm]
P_2&=k_{1}+p_{1}\quad&&\:P_5&&=k_{2}-p_{3}\quad&&\:P_8&&=k_{3}\quad&&\:N_1&&=k_{3}+p_{12}\quad&&\:N_4&&=k_{1}\\[-1mm]
P_3&=k_{1}+p_{12}\quad&&\:P_6&&=k_{2}-k_{3}\quad&&\:P_9&&=k_{3}+p_{1}\quad&&\:N_2&&=k_{2}+p_{1}\quad&&\:N_5&&=k_{2}\,.
\end{alignat*}
The definition of family $I_2$ is
\begin{alignat*}{10}
P_1&=k_{1}-k_{3}\quad&&\:P_4&&=k_{2}+p_{13}\quad&&\:P_7&&=k_{1}-k_{2}\quad&&\:P_{10}&&=k_{2}-k_{3}+p_{123}\quad&&\:N_3&&=k_{1}+p_{123}\\[-1mm]
P_2&=k_{1}+p_{3}\quad&&\:P_5&&=k_{2}+p_{123}\quad&&\:P_8&&=k_{3}\quad&&\:N_1&&=k_{3}+p_{13}\quad&&\:N_4&&=k_{1}\\[-1mm]
P_3&=k_{1}+p_{13}\quad&&\:P_6&&=k_{2}-k_{3}\quad&&\:P_9&&=k_{3}+p_{3}\quad&&\:N_2&&=k_{2}+p_{3}\quad&&\:N_5&&=k_{2}\,.
\end{alignat*}
}
\section{Matching boundary equations}\label{app: B}
In section~\ref{sec: boundaries}, we described a procedure for the analytic determination of boundary constants of a canonical differential equation system. 
The method requires matching solutions along paths that meet at the intersection of singular regions. 
In this appendix, we elaborate on the definition of a regularised limit for obtaining a boundary condition at a kinematic point with logarithmic singularity, where an ambiguity might arise. 
We illustrate the problem and its solution on a simplified example. 

Assume that we have the integral $J$ which depends on the kinematic invariants $x,y$ and on $\e$.
Let $J$ have the closed-form representation
\begin{align}
J(x,y;\e)= (x + 2 y)^{\e} b_0\,,
\label{eqs:diff_eq_example}
\end{align}
where $b_0$ is a boundary value. 
This integral satisfies the canonical differential equation 
\begin{align}
dJ(x,y;\e) = \e \, d\log(x+2 y)J(x,y;\e)\,
\label{eqs:diffexample}
\end{align}
and the origin is a singular point of the differential equation.
Now suppose that while fixing boundary conditions, we approach the origin along two paths,
\begin{align}
\mathcal{P}_1: (x,y) &= (\delta_1, m \, t_1^n)\,, \nonumber \\
\mathcal{P}_2: (x,y) &= (t_2, \delta_2)\,,
\end{align} 
where $m$, $n$ are non-zero real numbers, the parameters $t_i$ range from 0 to 1 and $\delta_i$ act as regulators.
Let us investigate the effect of the choice of parametrisation $y = m t_1^n$ on the matching of the solutions along these two paths as they approach the origin.

In the limit $\delta \rightarrow 0$, the differential equation~\eqref{eqs:diffexample} becomes
\begin{align}
\mathcal{P}_1&: \,  dJ(\delta_1 \sim 0, t_1 ;\e)=  dJ_1(t_1 ;\e) = \e \, d \log(m \,  t^n_1) J_1( t_1;\e )= \e \, n \,d \log(t_1)f_1( t_1;\e)\,,\nonumber \\
\mathcal{P}_2&: \, dJ(\delta_2 \sim 0, t_2;\e)= dJ_2( t_2;\e) = \e\,  d \log(t_2) J_2( t_2;\e)\,,
\label{eqs:examplelimit}
\end{align}
as $d\log(m)=0$.
Consequently, the solutions for the integral $J$, order by order in $\e$, along the two paths are
\begin{alignat}{6}
\mathcal{P}_1&: \, J_1(t_1;\e) &&= c^{(0)}_1 &&+ \e\bigl[n \log(t_1) &&+ c^{(1)}_1 \bigl]&&+ \mathcal{O}(\e^2)\,,\nonumber \\
\mathcal{P}_2&:  \, J_2(t_2;\e) &&= c^{(0)}_2 &&+ \e\bigl[ \log(t_2) &&+ c^{(1)}_2 \bigl]&&+ \mathcal{O}(\e^2)\,.
\label{eqs:sol}
\end{alignat}
Next, we need to approach the singular point $(x,y) =(0,0)$ by taking the limit $t_{i} \rightarrow 0$ and equate the two solutions, but it is unclear how to treat the divergent logarithms.

If we were to ignore the logarithmic divergences in eq.~\eqref{eqs:sol}, the matching would give
\begin{align}
c_1^{(0)} = c_2^{(0)}, \nonumber \\ 
c_1^{(1)} = c_2^{(1)}.
\label{eqs:matching wrong}
\end{align}
This corresponds to taking the following limit of the closed form~\eqref{eqs:diff_eq_example}, parametrised along the two paths,
\begin{align}
 \mathcal{P}_1&:  \lim_{t_1 \rightarrow 0}  \lim_{\delta_1 \rightarrow 0} t_1^{-n \,\e} J(\delta_1,m t^n_1;\e) = m^{\e} b_0\,,\nonumber \\
 \mathcal{P}_2&:  \lim_{t_2 \rightarrow 0}  \lim_{\delta_2 \rightarrow 0} t_2^{- \,\e} J(\delta_2, t_2;\e) = b_0\,,
\end{align}
which implies that the correct values of the constants are in fact
\begin{alignat}{4}
 \mathcal{P}_1&: \, c_1^{(0)} &&= b_0\,,\quad c_1^{(1)} &&= \log(m) b_0\,, \nonumber \\
 \mathcal{P}_2&: \, c_2^{(0)} &&= b_0\,, \quad c_2^{(n)} &&=0  \quad\text{for} \: n \geq1\,.
\end{alignat}
Clearly, the naive matching result~\eqref{eqs:matching wrong} omits powers of the constant $\log(m)$ at every $\e$-order.
Owing to the properties of the logarithm, the matching is insensitive to the rescaling of the parameter by a power $n$, but the multiplicative factor $m$ can introduce an ambiguity. 

The correct and unambiguous definition of the regularized value must therefore take into account any such factor.
To this end, we can rescale the integral by the factor which appears in the differential equation~\eqref{eqs:diff_eq_example} and thereby define the regularised value
\begin{align}
R(\e) \equiv \lim_{x,y \rightarrow 0} \tilde{J}(x,y) = \lim_{x,y \rightarrow 0}(x + 2 y)^{-\e}J(x, y;\e)\,.
\end{align}
Note that the function $\tilde{J}$ is now finite and continuous everywhere, including in the neighbourhood of $(0,0)$. 
With this definition, the limit is path-independent. 

Analogously, when we are dealing with a coupled set of differential equations given by the matrix of logarithmic forms $\tilde{M}$ as in eq.~\eqref{eqs:canonical_eq}, we shall define
\begin{align}
\vec{R}(\e) \equiv \lim_{x,y \rightarrow x_0,y_0}e^{-\e\tilde{M}}\vec{J}(x, y;\e)\,,
\label{eq: reglimit}
\end{align}
where $(x_0, y_0)$ is a singular point. 

This represents a completely general solution to the problem, but it requires the knowledge of $\vec{J}(x,y;\e)$.
Equivalently, one can take the limit of the solution of the regularised differential equation with the simplified alphabet order by order in $\e$, as in eq.~\eqref{eqs:sol} above. 
If during the derivation of the differential equations on the particular segment,
no multiplicative factors are generated, the logarithmic divergences analogous to those in eq.~\eqref{eqs:sol} can be safely ignored during the matching, reproducing exactly the result of~\eqref{eq: reglimit} order by order in $\e$. 
In our calculation, we explicitly verified that with our choice of segments and parametrisations, such factors are not generated in
the $\delta \rightarrow 0$ expansions of any logarithmic kernels in the differential equations.
Finally, it is always possible to check the correctness of the matching procedure by transporting the boundary constants along a closed path formed by the segments and verifying that the  same value is recovered.
\section{Organisation of the Supplementary Material}\label{app: C}
In this appendix, we briefly outline the organisation of the supplementary material attached to this manuscript. All attached files are in \texttt{Mathematica} format. 
\begin{itemize}
    \item \texttt{Families/}: For each integral family in fig.~\ref{tab: integral families}, we provide the list of propagators and numerators (see also app.~\ref{app: A}), a canonical basis and the corresponding differential equations, and a regularised boundary vector at the point $(x,y)=(0,0)$, obtained as described in sec.~\ref{sec: boundaries}.
    \item \texttt{BoundariesDemo/}: An executable notebook implementing the fixing of boundary constants using the method of sec.~\ref{sec: boundaries} for a two-loop example.
    \item \texttt{TridentFunctions/}: For each relevant canonical integral and kinematic crossing listed in \texttt{basis\_labels.m}, we provide a map to the trident functions $\psi_{1-1282}$ in the form of a sparse rational array of size $13812\times1282$ in \texttt{mapping\_positions.m} and \texttt{mapping\_values.m}.
    We also provide the UT Laurent expansions of the trident functions from $\e^0$ to the last relevant order $\e^6$ in terms of 42258 iterated integrals and transcendental constants in \texttt{sb\_labels.m}, again as a sparse array in \texttt{expansion\_positions.m} and \texttt{expansion\_values.m}.
    \item \texttt{FormFactor/}: We provide the three-loop finite remainders $R^{(3)}$ and $\mathcal{E}^{(3)}$ of the $\mathcal{N}=4$ sYM form factor in terms of Chen iterated integrals and MPLs.
    \item \texttt{Hjet/}: The three independent helicity coefficients for the decay of a $H$ boson to three partons at three loops are given in terms of iterated integrals and in a manifestly real MPL representation. In particular, we provide the part of the finite remainders containing new letters, with rational coefficients evaluated at $x=3/13,y=11/17$.
\end{itemize}

\bibliographystyle{JHEP}
\bibliography{main}
\end{document}